\documentclass[aps,prd,preprint,superscriptaddress,floatfix,showpacs]{revtex4-1}
\usepackage[english]{babel}
\usepackage[utf8]{inputenc}
\usepackage[T1]{fontenc}
\usepackage{amsmath,amssymb,braket,slashed,mathrsfs}
\usepackage{pifont}
\allowdisplaybreaks

\usepackage{graphicx}
\graphicspath{{./figures/}}
\usepackage{color,hyperref}
\usepackage{cleveref}

\crefformat{section}{Sec.~#2#1#3}
\crefmultiformat{section}{Sec.~#2#1#3}
    { and~#2#1#3}{, #2#1#3}{ and~#2#1#3}
\crefrangeformat{section}{Sec.~#3#1#4\,--\,#5#2#6}
\crefformat{subsection}{Sec.~#2#1#3}
\crefmultiformat{subsection}{Sec.~#2#1#3}
    { and~#2#1#3}{, #2#1#3}{ and~#2#1#3}
\crefrangeformat{subsection}{Sec.~#3#1#4\,--\,#5#2#6}
\crefformat{subsubsection}{Sec.~#2#1#3}
\crefmultiformat{subsubsection}{Sec.~#2#1#3}
    { and~#2#1#3}{, #2#1#3}{ and~#2#1#3}
\crefrangeformat{subsubsection}{Sec.~#3#1#4\,--\,#5#2#6}
\crefformat{figure}{Fig.~#2#1#3}
\crefmultiformat{figure}{Figs.~#2#1#3}
    { and~#2#1#3}{, #2#1#3}{ and~#2#1#3}
\crefrangeformat{figure}{Figs.~#3#1#4\,--\,#5#2#6}
\crefformat{table}{Table~#2#1#3}
\crefmultiformat{table}{Tabs.~#2#1#3}
    { and~#2#1#3}{, #2#1#3}{ and~#2#1#3}
\crefrangeformat{table}{Tabs.~#3#1#4\,--\,#5#2#6}
\crefformat{equation}{Eq.~(#2#1#3)}
\crefmultiformat{equation}{Eqs.~(#2#1#3)}{ and~(#2#1#3)}{, (#2#1#3)}{ and~(#2#1#3)}
\crefrangeformat{equation}{Eqs.~(#3#1#4)--(#5#2#6)}

\definecolor{linkcolor}{rgb}{.17578125,.1875,.5703125}
\hypersetup{
    pdfstartview={FitH},                         
    pdftitle={Electromagnetic finite-size effects to the hadronic vacuum polarization},
    pdfauthor={J. Bijnens, J. Harrison, N. Hermansson-Truedsson, T. Janowski, A. Jüttner, A. Portelli},                         
    pdfsubject={Physics article},                
    pdfcreator={Antonin Portelli},                               
    colorlinks=true,                             
    linkcolor=linkcolor,                         
    citecolor=linkcolor,                         
    filecolor=black,                             
    urlcolor=linkcolor      						 
}

\newcommand{\cf}{\textit{cf.}~}

\newcommand{\ie}{\textit{i.e.}~}

\renewcommand{\bar}{\overline}
\DeclareMathOperator{\bigo}{O}

\renewcommand{\epsilon}{\varepsilon}

\renewcommand{\tilde}{\widetilde}

\newcommand{\nsp}{\mathbf{n}}

\newcommand{\ksp}{\mathbf{k}}
\newcommand{\ellsp}{\boldsymbol{\ell}}

\newcommand{\vel}{\mathbf{v}(\boldsymbol{\ell})}

\newcommand{\vdku}{\vel\cdot\hat{\ksp}}
\newcommand{\qedl}{\mathrm{QED}_{\mathrm{L}}}
\newcommand{\uoe}{School of Physics and Astronomy, 
University of Edinburgh, Edinburgh EH9 3FD, United Kingdom}
\newcommand{\uos}{School of Physics and Astronomy, 
University of Southampton, Southampton SO17 1BJ, United Kingdom}
\newcommand{\lund}{Department of Astronomy and Theoretical Physics,\\
 Lund University, 223 62 Lund, Sweden}

\begin{document}
  \begin{flushright}
   LU TP 18-30\\
   March 2019 \\
   \end{flushright}

  \title{Electromagnetic finite-size effects\\ to the hadronic vacuum polarization}
  \author{J. Bijnens}\affiliation{\lund}
  \author{J. Harrison}\affiliation{\uos}
  \author{N. Hermansson-Truedsson}\affiliation{\lund}
  \author{T. Janowski}\affiliation{\uoe}
  \author{A. Jüttner}\affiliation{\uos}
  \author{A. Portelli}\email[corresponding author, ]{antonin.portelli@ed.ac.uk}
  \affiliation{\uoe}
  \begin{abstract}
    In order to reduce the current hadronic uncertainties in the theory prediction for the anomalous
    magnetic moment of the muon, lattice calculations need to reach sub-percent
    accuracy on the hadronic-vacuum-polarization contribution. This requires the  
    inclusion of $\mathcal{O}(\alpha)$ electromagnetic corrections. The
    inclusion of electromagnetic interactions in lattice simulations is known to
    generate potentially large finite-size effects suppressed only by  powers of the
    inverse spatial extent. In this paper we derive an analytic expression for
    the $\qedl$ finite-volume corrections to the two-pion contribution to the
    hadronic vacuum polarization at next-to-leading order in the electromagnetic
    coupling in scalar QED. The leading term is found to be of order $1/L^{3}$
    where $L$ is the spatial extent. A
    $1/L^{2}$ term is absent since the current is neutral and a photon
    far away thus sees no charge and we show that this result is universal. Our analytical
    results agree with results from the numerical evaluation of loop integrals as well as 
    simulations of lattice scalar $U(1)$
    gauge theory with stochastically generated photon fields. In the latter case the agreement is 
    up to exponentially suppressed finite-volume effects. For completeness we also calculate the
    hadronic vacuum polarization in infinite volume using a basis of 2-loop
    master integrals.
  \end{abstract}
  \maketitle
  \tableofcontents
  \newpage

\section{Introduction}

One of the most precisely measured quantities in particle physics is the
anomalous magnetic moment of the muon $a_\mu = \frac{g_\mu-2}{2}$, where $g_\mu$
describes the ratio of couplings of the muon spin and orbital angular momentum
to an external magnetic field. Historically, Dirac's original tree-level
prediction $g=2$ was in good agreement with experimental results, but the
discrepancy which eventually arose became very strong evidence in support of
quantum electrodynamics (QED). Both experimental measurement and Standard-Model
predictions for $a_\mu$ have by now reached a precision of about 0.5ppm where a
tension of 3.5-4$\sigma$ is observed~\citep{Bennett:2006fi,Jegerlehner:2017lbd, Davier:2017zfy,
Keshavarzi:2018mgv}. Efforts are therefore under way to increase the accuracy of
both measurements and theoretical predictions. To address the former two new
experiments have been planned, E989 at Fermilab~\citep{Logashenko:2015xab} and
E34 at J-PARC~\citep{Otani:2015lra}. The Fermilab experiment is expected to lead
to first new results in 2019 and the J-PARC experiment is expected to
begin in 2020. Both experiments aim at increasing the experimental accuracy by a
factor of 4 to 0.14ppm. To address the latter, we note that the main challenge
on the theoretical side comes from non-perturbative contributions, namely the
hadronic vacuum polarization (HVP) and hadronic light-by-light scattering
(HLbL), of which the HVP constitutes the dominant contribution to the
theoretical uncertainty. The traditional approach to estimating the HVP uses
dispersion relations together with the optical theorem to relate it to the
measured cross section of $e^+e^-$ to hadrons \citep{Davier:2017zfy,
Keshavarzi:2018mgv,Davier:2010nc, Hagiwara:2011af}. More recently, there has
been a significant progress in calculation of the HVP from first principles
using lattice QCD~\citep{Boyle:2011hu, DellaMorte:2011aa, Burger:2013jya,
Chakraborty:2014mwa, Chakraborty:2015cso, Bali:2015msa, Chakraborty:2015ugp,
Blum:2015you, Blum:2016xpd, Chakraborty:2016mwy, DellaMorte:2017dyu,
Borsanyi:2017zdw, Blum:2018mom, Giusti:2018mdh,Davies:2019efs}.

Based on simple power counting we expect strong and electromagnetic isospin
breaking effects to contribute at the percent level. Given that this corresponds
to the level of precision state-of-the-art lattice simulations are able to
achieve, these effects need to be included in future calculations. Here we
concentrate on electromagnetic effects which can be computed in the
lattice-discretized finite-volume theory in several ways. Common to all
approaches is the difficulty  of defining charged states in a finite volume with
periodic boundary conditions and the resulting singularities from photon
zero-modes which need to be dealt with. In QED$_{\rm
TL}$~\citep{Duncan:1996xy,Duncan:1996be,Borsanyi:2013lga,Ishikawa:2012ix,
Aoki:2012st,deDivitiis:2013xla,Fodor:2016bgu} the global zero mode is removed by
hand while in QED$_{\rm
L}$~\citep{Hayakawa:2008an,Davoudi:2014qua,Fodor:2015pna,Blum:2007cy,
Blum:2010ym,Ishikawa:2012ix,Giusti:2017dwk,Giusti:2017dmp,Giusti:2017jof,
Blum:2018mom,Boyle:2017gzv,Lee:2015rua,Matzelle:2017qsw,Davoudi:2018qpl} the
photon zero-mode is subtracted individually on every time slice. An alternative
avenue is to perform simulations with a massive photon~\citep{Endres:2015gda}
followed by an extrapolation to zero photon mass to obtain physical
results~\citep{Endres:2015gda, Bussone:2017xkb}. In yet another approach one
introduces charge-conjugation boundary
conditions~\citep{Polley:1993bn,Wiese:1991ku,
Kronfeld:1992ae,Kronfeld:1990qu,Lucini:2015hfa,Hansen:2018zre}  which allow for
constructing gauge-invariant charged states in a finite volume. QED corrections
have been performed in~\citep{Blum:2007cy, Blum:2010ym, Borsanyi:2013lga,
Borsanyi:2014jba, Horsley:2015vla, Horsley:2015eaa, Basak:2016jnn,
Fodor:2016bgu,Blum:2018mom}. Isospin-breaking corrections to the HVP have been
explicitly considered in~\citep{Giusti:2017jof, Boyle:2017gzv,
Chakraborty:2017tqp,Chakraborty:2018iyb,Blum:2018mom}.

A recurring systematic in QCD+QED calculations is the presence of large finite
volume (FV) effects, which scale as $1/L^n$ with the box size $L$ for some
exponent $n$. This is the result of the photon being a massless particle and the
long-ranged nature of electromagnetic interactions. The finite-volume
corrections have been studied in effective field theories for the meson masses
in \citep{Hayakawa:2008an,Borsanyi:2014jba, Davoudi:2014qua, Endres:2015gda,
Lucini:2015hfa,Davoudi:2018qpl} and decay rates in \citep{Lubicz:2016xro}. The
finite-volume correction to the HVP at order $\alpha$ however has not been
previously calculated and this is the subject of this paper. We will extend the
methodology for computing finite volume effects described in
\citep{Davoudi:2018qpl} to the electromagnetic correction to the HVP.

This paper is organized as follows. \cref{sec:hvp} describes the preliminaries
of the HVP function, which are relevant in both finite and infinite volumes.
\cref{sec:fv} describes the analytic derivation of the finite-volume correction
of the HVP with $\mathcal{O}(\alpha)$ electromagnetic correction. This is the
main result of this paper. Finally, \cref{sec:num} contains numerical tests of
the analytic expressions derived in \cref{sec:fv}. In~\cref{sec:infv} we present
a calculation of the NLO HVP in continuum Minkowski space.

\section{The hadronic vacuum polarization}\label{sec:hvp}
The main object of interest is the Euclidean 2-point function
\begin{align}\label{eq:hvpcurrent}
 \Pi _{\mu\nu}(q)  =  \, \int d^{4}x \, e^{iq\cdot x}\bra{0} 
 \mathrm{T}[j_{\mu}(x)j_{\nu}^{ \dagger}(0)]\ket{0}\, , 
\end{align}
where $j_{\mu} (x)$ is a charged or neutral vector current and $q^{2}$ is the
external, Euclidean photon momentum. We start by presenting the calculation for
neutral currents relevant for the HVP and then, since the calculation is
equivalent up to numerical factors in the summation of diagrams, briefly present
the result for charged ones. Note that for neutral currents Ward-Takahashi
identities imply that $\Pi _{\mu\nu}(q)= \left(q _\mu q_\nu-q^2 \delta
_{\mu\nu}\right) \Pi (q^2)$. The quantity $\Pi(q ^{2}) $ is ultraviolet
divergent and it is conventional to calculate the finite, subtracted quantity
\begin{align}
\hat{\Pi }  \left( q^{2} \right) = \Pi \left( q^{2}\right) - \Pi \left( 0 \right)  \, .
\end{align} 
This may be expanded in powers of the electric charge as $\hat{\Pi } \left(
q^{2} \right) = \hat{\Pi } ^{(0)} \left( q^{2} \right) + \hat{\Pi } ^{(1)}
\left( q^{2} \right) +\ldots $, where $ \hat{\Pi } ^{ (0)} \left( q^{2} \right)$
and $ \hat{\Pi } ^{ (1)} \left( q^{2} \right)$  are the leading order (LO) and
next-to leading order (NLO) terms, i.e. $\mathcal{O}\left(1\right) $ and
$\mathcal{O}\left( \alpha \right) $,  respectively.

\begin{figure}
  \centerline{
  \includegraphics[width=0.9\linewidth]{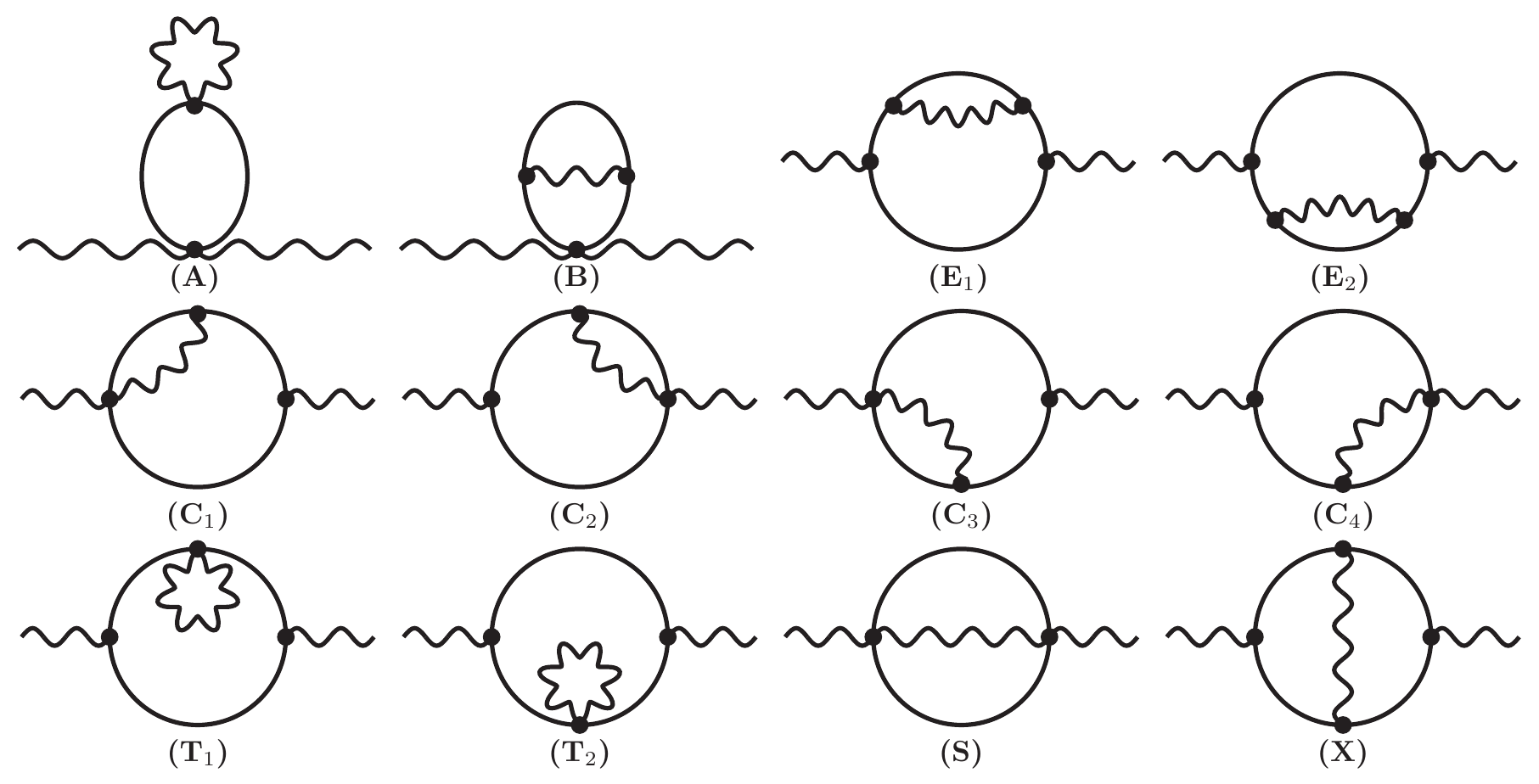}}
  \caption{The twelve connected diagrams contributing at NLO.}
  \label{fig:DiagramFig}
\end{figure}
\begin{figure}
\centerline{
\includegraphics[width=0.75\linewidth]{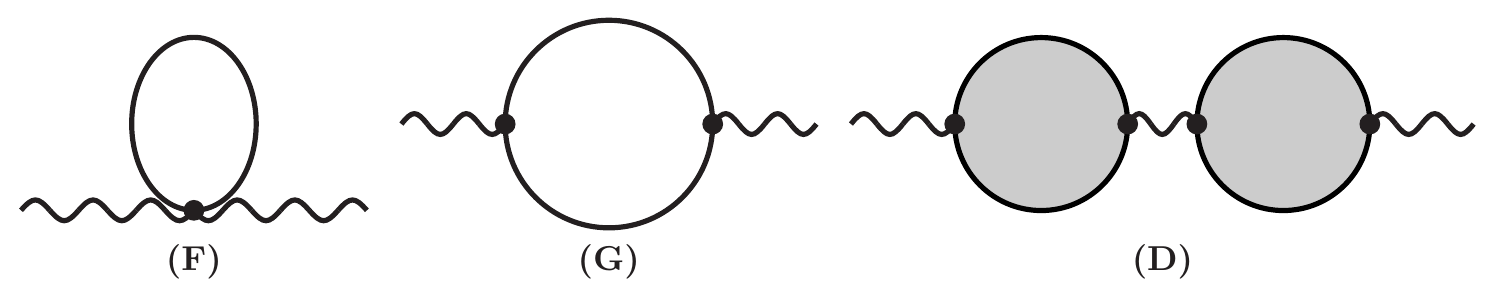}}
\caption{Diagrams (F) and (G) are the LO connected contributions to the HVP, whereas (D) is the NLO disconnected contribution consisting of four diagrams.}
\label{fig:DiagramLOFig}
\end{figure}

Although the two-pion contribution to the HVP is small compared to the vector
resonance one, it is the lightest contribution and dominates finite-volume
effects~\citep{Aubin:2015rzx}. Here we consider the electromagnetic corrections
to this contribution, and use scalar QED as an effective theory of elementary
pions. The scalar QED Lagrangian in Euclidean space is given by
\begin{align}
\mathcal{L} = \left(\partial_\mu\phi^*+i e A_\mu\phi^*\right)
\left(\partial_\mu\phi-i e A_\mu\phi\right)+m^2\phi^*\phi+\frac{1}{4}F_{\mu\nu}F_{\mu\nu} \, ,
\label{eq:LagrangianEuclEqn}\end{align}
for a scalar field $\phi$, a photon field $A_{\mu}$ and the electromagnetic
tensor $F_{\mu\nu} = \partial _{\mu}A_{\nu}-\partial _{\nu} A_{\mu}$. We only
consider the leading order scalar interactions, higher-order $\mathcal{O}\left( \lambda
\alpha^{2}\right)$ contributions, where $\lambda$ is the four-scalar vertex
coupling, enter at three loop order. The connected diagrams needed at NLO
for the HVP are therefore those in~\cref{fig:DiagramFig}. Seeing that some of
these diagrams are equal up to relabelling of momenta in the loops, we only need
to calculate seven topologies, namely (A), (B), (E$_{1}$), (C$_{1}$), (T$_{1}$),
(S) and (X). Diagrams (A) and (B) do not depend on the external momentum and
thus cancel in the subtraction, so only (E), (C), (T), (S) and (X) contribute.
The topology subscripts have here been suppressed and will remain so in
the rest of the paper. The  labelling refers to embedded sunrise (E),
contact (C), embedded tadpole (T), sunset (S) and photon exchange (X). It should
be noted that also the diagrams in (D) in~\cref{fig:DiagramLOFig} are in general
needed at NLO, but they are excluded for $\mathbf{q}=0$ in $\qedl$ and are in
infinite volume simply related to the square of the LO contribution, given by
diagrams (F) and (G), as will be discussed in more detail later. The specific
choice of the photon rest frame is elaborated on in~\cref{sec:fv}.

For completeness, we also calculated the NLO HVP in infinite volume. We considered both a Euclidean lattice using lattice perturbation theory, as well as continuum Minkowski space. The Minkowski space calculation is presented in~\cref{sec:infv}.

Using the Feynman rules from the continuum Euclidean space Lagrangian yields the momentum-space integrands of the diagrams
\begin{align}
(F) : \, \, & \frac{-2\delta _{\mu \nu }}{\ell^{2}+m^{2}},
  \label{eq:intF}\\
(G) : \, \, &\frac{\left( q-2\ell\right)_{\mu} \left(q- 2\ell\right) _{\nu}  }{\left( \ell^{2}+m^{2}\right) \left( \left( \ell-q\right) ^{2}+m^{2}\right)  },
  \label{eq:intG}\\
(A) : \,  \, &\frac{2d\, \delta _{\mu \nu}}{k^{2}\left( \ell^{2} + m^{2}\right) ^{2}},  \label{eq:intA}\\
(B) : \, \, &\frac{-2 \delta _{\mu \nu}(2\ell+k)^{2}}{k^{2}\left( \ell^{2}+m^{2}\right) ^{2}\left( \left( k+\ell\right) ^{2} + m^{2} \right)  },
  \label{eq:intB}\\
(E) : \, \, & \frac{\left( 2\ell-q\right) _{\mu}\left( 2\ell-q\right) _{\nu}(2\ell+k)^{2}}{k^{2}\left( \ell^{2}+m^{2}\right) ^{2}\left( \left( k+\ell\right) ^{2}+m^{2}\right)\left( \left( \ell-q\right) ^{2}+m^{2}\right)  },
  \label{eq:intE}\\
(C) : \, \, &\frac{-2\left(2q- 2\ell-k\right) _{\mu}\left( q-2\ell\right) _{\nu}  }{k^{2}\left( \ell^{2}+m^{2}\right) \left( \left( k+\ell-q\right) ^{2}+m^{2}\right) \left( \left( \ell-q\right) ^{2}+m^{2}\right)   },
    \label{eq:intC}\\
(T) : \, \, &\frac{-d\left( 2\ell-q\right) _{\mu} \left( 2\ell-q\right) _{\nu}}{k^{2}\left( \ell^{2}+m^{2}\right) ^{2}\left( \left( \ell-q\right) ^{2} +m^{2}\right) },
  \label{eq:intT}\\
(S) : \, \, & \frac{4\delta _{\mu \nu}}{k^{2}\left( \ell^{2}+m^{2}\right) \left( \left( k+\ell-q\right) ^{2}+m^{2}\right)  },
  \label{eq:intS}\\ 
(X) : \, \, & \frac{-\left(q- 2\ell\right) _{\mu} \left(q- 2\ell-2k\right) _{\nu} \left( 2q-2\ell-k\right) \cdot \left( 2\ell+k\right)  }{k^{2}\left( \ell^{2}+m^{2}\right)\left( \left( k+\ell\right) ^{2}+m^{2}\right) \left( \left( \ell-q \right) ^{2} +m^{2} \right) \left( \left( k+\ell-q\right) ^{2}+m^{2}\right)    },\label{eq:intX}
  \end{align}
where $k$ is the photon loop momentum, $\ell$ is the pion-loop momentum and $d$
is the number of dimensions. Similar expressions for Minkowski space are given
in~\cref{sec:MinkDiagApp}.

\section{Finite-size effects to the scalar vacuum polarization}\label{sec:fv}
We can express the renormalized HVP function $\hat{\Pi}(q^2)$ through the
following trace of the subtracted vector two-point function
\begin{align}\label{eq:PiFVEqn}
\hat{\Pi} \left( q^{2}\right) = \frac{1}{3q_{0}^{2}}\sum_{j=1}^3\left[ \Pi  _{jj}\left( q_{0},\mathbf{0}\right) -\Pi _{jj}\left( 0\right) \right] \, ,
\end{align}
where the photon rest frame has been specifically chosen. There are two main
reasons for choosing this frame. First and foremost, this is typically the frame
used in current lattice calculations. Moreover, it simplifies the finite-volume
calculation immensely, in particular as the coefficients $c_{j}$ defined below
then are independent of the photon momentum. 

Note that diagrams (A) and (B) automatically vanish in the subtraction
in~\cref{eq:PiFVEqn}, as they are independent of the external momentum.
Moreover, the disconnected contribution (D) is zero in $\text{QED}_{\mathrm{L}}$
because the photon propagator vanishes in the rest frame. We are thus left with
diagrams (E), (C), (T), (S) and (X), so that, including all permutations of the
diagrams, the $\mathcal{O}(\alpha)$ contribution to the HVP can be written as
\begin{align}\label{eq:combinatoricPiFV}
\Pi ^{(1)}\left( q^{2}\right)  =\, & 2\Pi_{E}(q^{2})+ 4\Pi_{C}(q^{2}) + 2\Pi_{T}(q^{2})+\Pi_{S}(q^{2}) + \Pi_{X}(q^{2})= \sum _{U} a_{U}\Pi _{U}(q^{2})\, ,
\end{align}
where $\Pi_{U}(q^2)$ denotes the contribution from diagram (U). Next we define
the corresponding integrand (excluding the factors of $2\pi$ in the measure) as
$\pi _{U}\left( k,\ell,q_{0}\right) $. 

In finite-volume, we assume space to be periodic with spatial extent $L$ and
time to remain infinite. We now present the procedure followed to determine the
finite-size effects to $\Pi^{(1)}(q^2)$ which decay like powers of $1/L$. This
strategy is a direct generalization of the procedure for one-loop integrals in
Ref.~\citep{Davoudi:2018qpl}. The remaining part of this section is a formal
description of our approach to compute the large volume expansion. Although the
final result presented in~\cref{sec:analyticFV_final} is quite compact,
intermediate expressions can be quite cumbersome. It is therefore desirable to
implement the whole strategy in a computer algebra system. The calculations
presented here were performed using {\sc FORM}~\citep{Vermaseren:2000nd} and
Mathematica~\citep{Mathematica}, and the associated Mathematica notebook is
provided as a supplement of this paper under the General Public License version
3. Intermediate products of the derivation are provided for future reference
in~\cref{sec:diagramapp}.

For a given diagram, we start by computing the two
energy integrals in $k_0$ and $\ell_0$ using contour integration. Feynman
integrands are rational functions and this integration is systematically
feasible analytically. We thus obtain
\begin{align}
\rho _{U}\left( \mathbf{k},\ellsp,q_{0}\right) =  \int \frac{dk_{0}}{2\pi} \frac{d\ell_{0}}{2\pi}\pi _{U}\left( k,\ell ,q_{0}\right) \, .
\end{align}
In analogy with~\cref{eq:PiFVEqn}, we also define the subtracted quantities
$\hat{\rho}_{U}$. The finite volume effects on $\hat{\Pi}(q^2)$ for diagram (U)
in $\text{QED}_{L}$ can then be written as
\begin{align}
\Delta \hat{\Pi} _{U}\left( q_{0}^{2}\right) = \left( \frac{1}{L^{6}}\left. \sum_{\mathbf{k}} \right.  ^{\prime }\sum _{\mathbf{\ellsp}}-\int \frac{d^{3}\mathbf{k}}{\left( 2\pi\right) ^{3}}\frac{d^{3}\ellsp}{\left( 2\pi\right) ^{3}}\right)  \hat{\rho} _{U}\left( \mathbf{k},\ellsp,q_{0}\right) \,,
\end{align}
where finite-volume sums are on quantized momenta of the form
$\ksp=\frac{2\pi}{L}\nsp$ with $\nsp$ a vector with integer components, and a
primed sum means that the origin is excluded, which here comes from the
$\mathrm{QED}_{\mathrm{L}}$ prescription. One important aspect here is that we
are only considering the $q^2>0$ case. This means that pions in diagrams are
purely virtual and cannot generate power-like finite-volume effects through
on-shell singularities. Using the Poisson summation formula for the pion part
yields
\begin{align}\label{eq:poissoneqn}
\Delta \hat{\Pi } _{U}\left( q_{0}^{2}\right) = \left( \frac{1}{L^{3}}\left. \sum_{\mathbf{k}} \right.  ^{\prime }-\int \frac{d^{3}\mathbf{k}}{\left( 2\pi\right) ^{3}}\right)  \int \frac{d^{3}\ellsp}{\left( 2\pi\right) ^{3}} \hat{\rho} _{U}\left( \mathbf{k},\ellsp,q_{0}\right) +\cdots \, ,
\end{align}
where the omitted terms denoted by ellipsis are the exponentially suppressed
contributions from the virtual pions. 

To determine power-like finite-size effects in the five diagrams (E), (C), (T), (S) and (X), we closely
follow the strategy laid out in~\cite{Davoudi:2018qpl}.
One starts by isolating the singularities in the photon momentum $\ksp$ in $\rho
_{U}(\mathbf{k},\ellsp,q_{0})$,
\begin{align}
\rho _{U}\left( \mathbf{k},\ellsp,q_{0}\right)  = \sum _{j=0}^{n_{U}}\left( \frac{2\pi }{\left| \mathbf{k}\right|}\right) ^{j}u_{j}\left( \hat{\mathbf{k}},\ellsp,q_{0}\right)  + \bar{\rho} _{U}\left( \mathbf{k},\ellsp,q_{0}\right) \, ,
\label{eq:rhoexp}
\end{align}
where $n_{U}$ is an integer that depends on the diagram in question,
$\hat{\mathbf{k}} = \mathbf{k}/\left| \mathbf{k}\right|$, and $\bar{\rho} _{U}(
\mathbf{k},\ellsp,q_{0})$ is an analytic function in the norm $\left| \mathbf{k}
\right|$ such that $\bar{\rho} _{U}( \mathbf{0},\ellsp,q_{0})=0$. The analytical structure of
all five diagrams is such that $n_U\leq 1$. If we now substitute
$\mathbf{k} = \frac{2\pi}{L} \mathbf{n}$ and expand in $1/L$, the finite volume
effects for diagram $U$ can be written as a power series in $1/L$ (up to exponentially small corrections),
\begin{align}
\Delta \hat{\Pi} _{U}\left( q_{0}^{2}\right) = \frac{\xi _{1}^{U}\left( q_{0}^{2}\right) }{L^{2}}+\frac{\xi _{0}^{U}\left( q_{0}^{2}\right) }{L^{3}}+\mathcal{O}\left(\frac{1}{L^4},e^{-mL}\right) \, .\label{eq:xiexp}
\end{align}  
The coefficients $\xi _{j}^{U} \left( q_{0}^{2} \right) $ are given by 
\begin{align}\label{eq:xiform}
\xi _{j}^{U} \left( q_{0}^{2}\right) =\Delta  _{\mathbf{n}}^{\prime } \left[ \frac{1}{\left| \mathbf{n}\right| ^{j}}\int \frac{d^{3}\ellsp}{\left( 2\pi\right) ^{3}}u_{j}\left( \hat{\mathbf{n}},\ellsp ,q_{0}\right) \right]\,.
\end{align} 
where $\Delta  _{\mathbf{n}}^{\prime }$ is, as in Ref.~\citep{Davoudi:2018qpl},
the $\mathrm{QED}_{\mathrm{L}}$ sum-integral difference operator
\begin{equation}
    \Delta  _{\mathbf{n}}^{\prime }=\left.\sum_{\mathbf{n}}\right.^{\prime }-\int d^{3}\mathbf{n}\,.
\end{equation}
Although $\bar{\rho} _{U}( \mathbf{k},\ellsp,q_{0})$ is an analytic function in
the norm $\left| \mathbf{k} \right|$, the norm itself in not analytic in the
components of $\ksp$ at the origin, which generates the $\mathcal{O}(1/L^4)$
effects in~\cref{eq:xiexp}.  We will now present the full expressions for the
finite-size effects to each of the five diagram topologies (S), (T), (C), (E)
and (X).

\subsection{The full finite-size effects}
\label{sec:analyticFV_final}
To find the finite-size effect to a diagram (U), the last step to perform is the
calculation of the $\xi_{j}^{U}$ coefficients in~\cref{eq:xiform}. For the
specific kinematics chosen here, i.e., spatial momenta equal to zero  (\cf\cref{eq:PiFVEqn}), the integrand
$u_{j}\left( \hat{\mathbf{n}},\ellsp ,q_{0}\right)$ is independent of the photon
momentum direction $\hat{\mathbf{n}}$. The function $\xi_{j}^{U}(q_0^2)$ then has the
form
\begin{equation}
    \xi_{j}^{U}(q_0^2)=c_j\phi_j(q_0^2)\,,\label{eq:xijcj}
\end{equation} 
where identically to Ref.~\citep{Davoudi:2018qpl}, we define the coefficients
$c_{j} =\Delta  _{\mathbf{n}}^{\prime }|\mathbf{n}|^{-j}$. These can be
calculated numerically in several ways, and one possibility is presented
in~\citep{Davoudi:2018qpl}. The first three coefficients are $c_{0}=-1$, $c_{1}
= -2.83729748\ldots $ and $c_{2}= \pi c_{1}$.

The functions $\phi_j(q_0^2)$ in~\cref{eq:xijcj} can be
written as linear combinations of integrals of the form 
\begin{equation}
\Omega_{\alpha,\beta}(z)=\frac{1}{2\pi^2}\int_0^{\infty}d x\,
  x^2\omega_{\alpha,\beta}(x,z) \, , 
\end{equation}
where $z=q_0^2/m^2$, and
\begin{equation}
\omega_{\alpha,\beta}(x,z)=
  \frac{1}{( x ^{2}  +1)^{\frac{\alpha}{2}}[z+4(x ^{2}+1)]^{\beta}}\,.
\end{equation}
They arise after integrating the angular dependence of the integrals over
momentum $\ellsp$ for which $\left| \ellsp \right| =mx $. There are several
useful recursion relations and properties for these integrals, which we summarise
in~\cref{sec:OmegaApp}.  For instance, for $\alpha + 2\beta >3$ in $d=4$,
it is possible to write any $\Omega _{\alpha ,
\beta}$  as a linear combination of the six simple
functions $\Omega _{2,1}$, $\Omega _{3,1}$, $\Omega _{4,1}$, $\Omega _{5,1}$,
$\Omega _{0,2}$ and $\Omega _{1,2}$ as well as their respective derivatives. The
complete list of expressions that lead to these integrals, in particular, the
expansion~\cref{eq:rhoexp}, are given explicitly for all diagram topologies
in~\cref{sec:diagramapp}.

Finally, we summarize here the final expressions for the finite-volume effects
to each diagram, where every $\Omega_{\alpha,\beta}$ term implicitly depends on
$z=q_0^2/m^2$
\begin{align}\label{eq:hvpFV_diagrams}
\Delta \hat{\Pi}_E (z)=&\frac{c_{1}}{\pi m^{2}L^{2}} \left(
        - \frac{4}{3}\,\Omega _{-1,3}
        + \frac{1}{2}\,\Omega _{1,2}
        + \frac{4}{3}\,\Omega _{1,3}
        - \frac{1}{4}\,\Omega _{3,1}
        \right)
    \nonumber \\ &
        - \frac{c_{0}}{m^{3}L^{3}} \left(
        - \frac{8}{3}\,\Omega _{0,3}
        + \frac{32}{3}\,\Omega _{0,4}
        + \frac{1}{16}\,\Omega _{2,2}
        + \frac{10}{3}\,\Omega _{2,3}\right.\nonumber\\&
        \qquad\qquad\quad\left.
        - \frac{32}{3}\,\Omega _{2,4}
        - \frac{23}{128}\,\Omega _{4,1}
        + \frac{5}{16}\,\Omega _{4,2}
        - \frac{2}{3}\,\Omega _{4,3}
        \right)\,, \\  
\Delta \hat{\Pi}_C (z)=& \frac{c_{1}}{\pi m^{2}L^{2}} 
            \frac{1}{8}\,\Omega _{3,1}
        -   \frac{c_{0}}{m^{3}L^{3}} \left(
         \frac{8}{3}\,\Omega _{0,3}
        + \frac{1}{6}\,\Omega _{2,2}
        - \frac{8}{3}\,\Omega _{2,3}
        + \frac{1}{8}\,\Omega _{4,1}
        - \frac{1}{6}\,\Omega _{4,2}
        \right)\\
\Delta \hat{\Pi}_T(z)=& \frac{c_{1}}{\pi m^{2}L^{2}}  
            \frac{1}{4}\,\Omega _{3,1}\,,\\
\Delta \hat{\Pi}_S(z)=&-\frac{c_{1}}{\pi m^{2}L^{2}} 
            \frac{1}{4}\, \Omega _{3,1}
        +    \frac{c_{0}}{m^{3}L^{3}} \left(
         2\, \Omega _{2,2}
        + \frac{1}{4}\, \Omega _{4,1}
        \right)\,,\\
\Delta \hat{\Pi}_X(z)=&  \frac{c_{1}}{\pi m^{2}L^{2}} \left(
        \frac{8}{3}\,\Omega _{-1,3}
        - \Omega _{1,2}
        - \frac{8}{3}\,\Omega _{1,3}
        - \frac{1}{4}\,\Omega _{3,1}
        \right)
    \nonumber \\&- \frac{c_{0}}{m^{3}L^{3}} \left(
        - \frac{128}{3}\,\Omega _{-2,4}
        - \frac{16}{3}\,\Omega _{0,3}        
        + 64\, \Omega _{0,4}
        - \frac{11}{24}\,\Omega _{2,2}
        + \frac{20}{3}\,\Omega _{2,3}
        \right. \nonumber \\ & \left.\qquad\qquad\quad
        - \frac{64}{3}\,\Omega _{2,4}
        - \frac{17}{64}\,\Omega _{4,1} 
        + \frac{29}{24}\,\Omega _{4,2}
        - \frac{4}{3}\,\Omega _{4,3}
        \right)\,,
\end{align}
and where all the expressions are given up to
$\mathcal{O}(\frac{1}{L^4},e^{-mL})$ corrections. We can sum these terms
according to~\cref{eq:combinatoricPiFV}. The resulting series in $1/L$ for the
HVP at NLO is
\begin{equation}\label{eq:finalhvp}
\Delta \hat{\Pi} (q^2) =
\frac{c_0}{m^3 L^3}
\Bigg(
           \frac{16}{3}\Omega_{0,3}
          + \frac{5}{3}\Omega_{2,2}
          - \frac{40}{9}\Omega_{2,3}
          + \frac{3}{8}\Omega_{4,1}
          - \frac{7}{6}\Omega_{4,2}
          - \frac{8}{9}\Omega_{4,3}
\Bigg)\, ,
\end{equation}
where one notices the important cancellation of the $1/L^{2}$ terms. This result
can be understood from the underlying physics since the current is neutral and a
photon far away thus sees no charge. This cancellation has potentially important
consequences regarding the prediction of the contribution
$a_{\mu}^{\mathrm{HVP}}$ from the HVP to $a_{\mu}$ using lattice simulations.
Indeed, for typical physical simulations with $mL>4$, one has $1/(mL)^3<1.5\%$.
Under the safe assumption that the QED corrections to $a_{\mu}^{\mathrm{HVP}}$
are $\mathcal{O}(1\%)$, the electromagnetic finite-size effects discussed here
would represent a contribution smaller than $0.02\%$, well below the $0.1\%$
level required to reduce by a factor of $4$ the current theoretical
uncertainties on $a_{\mu}^{\mathrm{HVP}}$. Finally, for $mL>4$ one has
$e^{-mL}<1.8\%$, which means that in this regime the new, power-like finite-size
corrections introduced by QED are in principle not dominant compared to the
exponential QCD effects. In the following sections, we demonstrate that this
cancellation does not occur for charged currents, and that it is universal in
full QCD+QED and therefore directly applicable to lattice results.
\subsection{Charged currents}
For the neutral currents only charged pions are considered. If also $\pi ^{0}$
is included, the current $j_{\mu}$ in~\cref{eq:hvpcurrent} can be charged and
the current-current correlator can therefore be rewritten as
 \begin{equation}\label{eq:hvpcurrentcharged}
 \Pi_{\mu\nu}^{\textrm{charged}}(q )  =  \, \int d^{4}x \, e^{iq\cdot x}\bra{0}\mathrm{T}[j_{+\,  \mu}(x)j_{-\, \nu}(0)]\ket{0}  = \delta _{\mu \nu}q^{2}\Pi _{1}\left( q^{2}\right) -q_{\mu}q_{\nu}\Pi _{2}(q^{2})\, ,
\end{equation}
for two functions $\Pi _{1}(q^{2})$ and $\Pi _{2}(q^{2})$ that are equal for
neutral currents. We are again interested in the case
$q_{\mu}=(q_{0},\mathbf{0})$ and calculate the subtracted quantity
\begin{align}
\hat{\Pi } ^{\textrm{charged}}(q^{2})=\frac{1}{3q^{2}}\sum_{j=1}^3\left[ \Pi ^{\textrm{charged}}_{jj}(q_{0},\mathbf{0}) -\Pi ^{\textrm{charged}}_{jj}(0)\right]  = \Pi _{1}(q_{0}^{2}) -\Pi _{1}(0) \,.
\end{align}  
The function $\Pi _{1}(q^{2})$ can be expanded in the electromagnetic coupling
just as before, and we here denote the NLO contribution by $\Pi
_{1}^{(1)}(q^{2})$. The possible topologies of the NLO diagrams are the same
also here, but having charged currents implies that some of them may be
forbidden and the overall numerical factors can be different compared to the
netural case for those that are not. In order to find these differences, we
include the neutral pion by defining the meson matrix $M$ and the current matrix
$J_{\mu}$ as  
\begin{equation}
M=\left( 
\begin{array}{cc}
\frac{1}{\sqrt{2}}\pi ^{0} & \pi ^{+} \\
\pi ^{-}& -\frac{1}{\sqrt{2}} \pi ^{0}
\end{array}
\right) , \; \; \; \; J _{\mu}= \left( 
\begin{array}{cc}
\frac{2}{3} j_{\mu} & j _{+ \, \mu} \\
j_{-\, \mu}& -\frac{1}{3}j_{\mu} 
\end{array}
\right).
\end{equation}
The covariant derivative of $M$ can then be put in the form
\begin{equation}
D_{\mu}M = \partial _{\mu}M-i\left[ J_{\mu }, M\right] =  \left( 
\begin{array}{cc}
 \frac{1}{\sqrt{2}}\partial _{\mu }\pi ^{0}+i\pi ^{+}j_{-\, \mu }-i\pi ^{-}j_{+\, \mu } & \partial _{\mu}\pi ^{+}-i\pi ^{+}j_{\mu }+i\sqrt{2}\pi ^{0}j_{+\, \mu } 
\\
\partial _{\mu}\pi ^{-}+i\pi ^{-}j_{\mu }+i\sqrt{2}\pi ^{0}j_{-\, \mu } &  -\frac{1}{\sqrt{2}}\partial _{\mu }\pi ^{0}-i\pi ^{+}j_{-\, \mu }+i\pi ^{-}j_{+\, \mu }
\end{array}
\right),
\end{equation}
and the kinetic part of the Lagrangian is given by
\begin{align}
\mathcal{L}_{kin} =& \frac{1}{2}\textrm{tr}\left[ D_{\mu }M\left( D_{\mu }M\right) ^{\dagger } \right]  
= 
\frac{1}{2}\left( \partial _{\mu}\pi ^{0}\right) ^{2} + \partial _{\mu} \pi ^{+}\partial _{\mu }\pi ^{-} + i\sqrt{2}j_{+\, \mu}\left( \pi ^{0}\partial _{\mu}\pi ^{-}-\pi ^{-}\partial _{\mu} \pi ^{0}\right) 
+ \nonumber \\  
& +i\sqrt{2}j_{-\, \mu }\left( \pi ^{+}\partial _{\mu}\pi ^{0}-\pi ^{0}\partial _{\mu} \pi ^{+}\right)  +ij_{\mu}\left( \pi ^{-}\partial _{\mu}\pi ^{+}-\pi ^{+}\partial _{\mu} \pi ^{-}\right) -\sqrt{2}j_{\mu}j_{-\, \mu }\pi ^{+}\pi ^{0}
+ \nonumber \\ 
& -\sqrt{2}j_{+\, \mu}j_{ \mu }\pi ^{0}\pi ^{-} +j_{\mu}j_{ \mu }\pi ^{+}\pi ^{-}-j_{+\, \mu} j_{+\, \mu }\pi ^{-}\pi ^{-}-j_{-\, \mu}j_{-\, \mu }\pi ^{+}\pi ^{+} \nonumber \\ 
& +2j_{+\, \mu}j_{-\, \mu }\left( \pi ^{+}\pi ^{-}+\pi ^{0}\pi ^{0}\right) \, .
\end{align}
Using this, we find that diagram (X), one of the permutations of diagram (T),
two of the permutations of (C) and one of the permutations of diagram (E) are
forbidden. Moreover, the overall numerical factors change such that the relevant
NLO contribution becomes
\begin{align}\label{eq:combinatoricPiCharged}
\Pi _{1}^{(1)}\left( q^{2}\right)  =\, & 2\cdot (E) + 2\cdot (C) + 2(T)+\frac{1}{2}(S) .
\end{align}
This yields the NLO FV effects as
\begin{align}
          \Delta \hat{\Pi } ^{\textrm{charged}}\left( q^{2} \right) =&
           \frac{c_{1}}{\pi m^{2}L^{2} }\left(
          - \frac{8}{3} \Omega _{-1,3}
          + \Omega _{1,2}
          + \frac{8}{3} \Omega _{1,3}
          + \frac{1}{8} \Omega _{3,1}
          \right) 
\nonumber \\ &
  -\frac{c_{0}}{m^{3}L^{3}} \left(
          - \frac{13}{24} \Omega _{2,2}
          + \frac{20}{9} \Omega _{2,3}
          - \frac{15}{64} \Omega _{4,1}
          + \frac{7}{24} \Omega _{4,2}
          + \frac{4}{9} \Omega _{4,3}
          \right)\, ,
\end{align}
up to
$\mathcal{O}(\frac{1}{L^4},e^{-mL})$ corrections. The $1/L^{2}$ part does not vanish here. This is expected, since the current
no longer is neutral and the physical argument used for the neutral case no
longer applies.  
\subsection{Universality of the finite-size corrections}
In the above we showed that in one-loop scalar QED the leading contribution to
the FV effects on the subtracted vacuum polarization function $\hat{\Pi}(q^2)$
is of order $1/L^3$ in $\textrm{QED}_L$. We will now show that this conclusion
is independent of the effective-field-theory formulation chosen for the finite
volume calculation.

The $\mathcal{O}(\alpha)$ corrections to the current-current correlator
$\Pi_{\mu\nu}(q)$, which we denote $\Pi_{\mu\nu}^{(1)}(q)$, can be written as
\begin{equation}
    \Pi_{\mu\nu}^{(1)}(q) = \int \frac{d^4k}{{(2\pi)}^4} \int_{x,y,z} \bra{0}\mathrm{T}[j_\mu(x) j_\nu(y) j_\rho(z) j_\sigma(0)]\ket{0} e^{i q \cdot (x-y)} e^{i k \cdot z} \frac{\delta_{\rho\sigma}}{k^2}\,,
\end{equation}
where the symbol $\int_{x,y,z}$ is an abbreviation for the  integration
over the $3$ space-time positions $x$, $y$, and $z$. This amplitude is identical
to the amputated light-by-light scattering Green's function with two legs
contracted with the photon propagator. We will argue below that the
light-by-light matrix element at our choice of kinematics is free of
singularities and therefore expected to have only exponential finite volume
corrections. As a consequence, the only source of power-like corrections will
necessarily have to come from the photon propagator pole.

A useful form-factor decomposition of the light-by-light amplitude with form
factors that are free of kinematic singularities is given in equation (3.14)
of~\citep{Colangelo2015}. We note that all tensor structures have at least one
factor of each of the $q_1$, $q_2$, $q_3$ and $q_4$. In our formula, we need to
replace two of those external momenta with $k$ and the remaining two with $q$
and $-q$, respectively. Therefore all the tensor structures will be proportional to
either $k^2$ or $k_\mu k_\nu$ for some Lorentz indices $\mu$ and $\nu$, which in
turn can be replaced with $k^2$ using the $d$--dimensional integral (or sum in
FV) formula:
\begin{equation}
    \int \frac{d^dk}{{(2\pi)}^d} k_\mu k_\nu f(k^2) = \frac{\delta_{\mu\nu}}{d}\int \frac{d^dk}{{(2\pi)}^d} k^2 f(k^2)\,.
\end{equation}
This means that each of the tensor structures contribute a factor of $k^2$,
which cancels the pole in the photon propagator. Since the light-by-light form
factors $F(k^2,q^2,k\cdot q)$ are free of kinematic singularities, and we work
with Euclidean momenta which cannot give rise to any on-shell singularities,
they do not have any poles in $k$. This means that the leading FV correction
will come from a term which is constant in $k$, which has the contribution
proportional to $c_0/L^3=-1/L^3$, completing the proof.

The above argument can be simplified by considering the scalar HVP form factor
\begin{equation}
    \delta_{\mu\nu} \Pi^{(1)}_{\mu\nu}(q) = (d-1) q^2 \Pi^{(1)}(q^2) \, .
\end{equation}
The function $\Pi$ then has the form
\begin{align}
    \Pi^{(1)}(q^2) &= \frac{1}{(d-1)q^2} \int \frac{d^dk}{(2\pi)^d} K_{\rho\sigma}(k,q) \frac{\delta_{\rho\sigma}}{k^2} \, , \\
    K_{\rho\sigma}(k,Q) &= \int_{x,y,z}\bra{0} j_\mu(x) j_\mu(y) j_\rho(z)j_\sigma(0)\ket{0}e^{i q \cdot (x-y)} e^{ik\cdot z}
\end{align}
where the kernel $K_{\rho \sigma}$ satisfies the Ward identities $k_\rho
K_{\rho\sigma} = k_\sigma K_{\rho\sigma}=0$. The kernel function
$K_{\rho\sigma}$ is a momentum space amplitude and any singularities must
correspond to physical states. In Euclidean space with external momenta being
real, the function can not have any poles corresponding to physical states,
which would have to satisfy $p^2+m^2=0$ where $p$ is the momentum going through
any cut in the diagram. We can decompose $K_{\rho\sigma}$ in a series of tensor
structures multiplying form factors, which can generally be  written as
\begin{equation}
    K_{\rho\sigma}(k,q) = \delta_{\rho\sigma} F_0(k^2,q^2,k\cdot q) + \sum_{p,\ell \in \{k,q\}} p_\rho \ell_\sigma F_{p\ell}(k^2,q^2,k\cdot q)\,.
\end{equation}
Since $K_{\rho\sigma}$ is free of singularities and the tensor structures are
linearly independent, the form factors must be free of singularities as well.
The Ward identities $k_\rho K_{\rho\sigma} = k_\sigma K_{\rho\sigma}=0$ impose a
relation on the form factors simplifying the expression to
\begin{align}
    K_{\rho\sigma}(k,q) &= (k_\rho k_\sigma - k^2 \delta_{\rho\sigma})F_{kk}(k^2,q^2,k\cdot q) \notag\\
    & \quad + \left[ -(k\cdot q) \delta_{\rho\sigma} + k_\rho q_\sigma + k_\sigma q_\rho - \frac{q_\rho q_\sigma k^2}{k\cdot q} \right]F_{kq}(k^2,q^2,k\cdot q)\,.
\end{align}
As noted before, the form factors $F_{kk}$ and $F_{kq}$ are free of
singularities, however $F_{kq}$ must have a zero at $k\cdot q=0$ to cancel the
pole originating from the tensor structure it is multiplying. We conclude that
$F_{kq}$ must be proportional to $k\cdot q$. Finally, the form factor
decomposition of $K$ consistent with Lorentz symmetry, parity, and Ward identities is
\begin{align}
    K_{\mu\nu}(k,Q) &= \left[ q_\mu q_\nu k^2 - k_\mu q_\nu (k \cdot q) - q_\mu k_\nu (k\cdot q) + \delta_{\mu\nu}(k \cdot q)^2 \right] 
    F_1(k^2,q^2,k\cdot q) \nonumber\\
               &\quad+ \left( k_\mu k_\nu - k^2 \delta_{\mu\nu} \right) F_2(k^2,q^2,k\cdot q) \, .
\end{align}
As before, we note that the form factors $F_1 = -F_{kq}/(k\cdot q)$ and $F_2
=F_{kk}$ do not have poles in $k^2$ and the tensor structure has two factors of
$k$ which become $k^2$ under the integral, which cancels the photon propagator
pole. As before, this results in the leading contribution to the FV correction
to be proportional to $c_0/L^3$.

\section{Numerical validation}\label{sec:num}
In this section we provide two different numerical checks of the finite-volume
corrections derived in~\cref{sec:analyticFV_final}. Scalar QED is ideally suited
for numerical simulations. Indeed, as we will now explain in detail,
the theory can be  written on a
discrete space-time simply by replacing derivatives with finite differences. In the two
following subsections, we describe two different Monte-Carlo strategies to compute the
volume dependence of the scalar vacuum polarization. Firstly, the master
formula~\cref{eq:poissoneqn} is evaluated directly using a Monte-Carlo
integrator. Secondly, the finite-volume vacuum polarization is calculated at
$\mathcal{O}(\alpha)$ using lattice -scalar-QED simulations, following the
strategy described in~\citep{Davoudi:2018qpl}. Finally, we  discuss the
comparison of these results with analytical predictions.
\subsection{Scalar QED on a lattice}\label{lptSec}
In this section we explain our definition of the lattice discretized theory. The
conventions and notations are identical to~\citep{Davoudi:2018qpl}. We consider
space-time to be an Euclidean four-dimensional lattice with spatial extent $L$,
time extent $T$, and lattice spacing $a$.  Lattice QED is then 
defined by the action
\begin{align}\label{eq:latticeAction}
S\left[ \phi, A\right] = S_{\phi} \left[ \phi , A\right] + S_{A}[A]  ,
\end{align}
with scalar and gauge actions
\begin{align}
 S_{\phi} \left[ \phi , A\right] =& \,\frac{a^{4}}{2}\sum _{x }\Bigg[\sum_\mu|D_{\mu}\phi(x)|^2+m^{2}_{0}\left| \phi (x) \right| ^{2}\Bigg] =\frac{a^{4}}{2}\sum _{x}\phi ^{*}(x)\Delta \phi (x) , \nonumber \\
 S_{A}[A] =&\, \frac{a^4}2\sum _{x,\mu} \left[\sum _{\nu}\frac 12F_{\mu \nu}(x)^2 + \left[\delta_\mu A_\mu(x)\right]^2\right]=-\frac {a^4}2\sum_{x,\mu} A_\mu(x)\delta^2 A_\mu(x)\, ,
\end{align}
respectively, with $\Delta = m^{2}-\sum _{\mu}D^{*}_{\mu}D_{\mu}$. The summation
is over all the sites of the lattice. The covariant derivative  is defined
in terms of the $U(1)$ gauge link $U_\mu(x)=e^{iqaA_{\mu}(x)}$, where $q$ is the
electric charge of the scalar particle, as
\begin{equation}
 D_{\mu} \phi(x) = \frac 1a\left[ U_{\mu}(x)\phi(x+a\hat{\mu})-\phi(x)\right]\,,\;\;
 D_{\mu}^{*} \phi = \frac 1a\left[ \phi(x) -U^{\dagger}(x-a\hat{\mu})\phi(x-a\hat{\mu})\right]\,.
 \label{eq:lcovd}
\end{equation}
We also introduce the forward derivative $\delta_\mu A_\mu(x) =
a^{-1}[A_\mu(x+a\hat{\mu})-A_\mu(x)]$, which appears in the Feynman gauge-fixing
term. The electromagnetic tensor is  defined as
\begin{align}
  F_{\mu\nu}(x) &= \delta_\mu A_\nu(x) - \delta_\nu A_\mu(x)\,.
\end{align}
Expectation values in this theory are expressed in terms of the path integral
\begin{align}\label{eq:QEDPI}
\langle O\rangle =&\frac 1{\mathcal{Z}_{\rm L}}\int \mathcal D A\,\mathcal{D} {\phi}\,\mathcal{D}{\phi^\ast}\,
		O[\phi,\phi^\ast] \,e^{-S_{\rm L}[\phi,A]}\,,
\end{align}
where the integral measures represent integrations over the field variable at
each lattice site. The subscript L indicates that we are working within the
QED$_{\rm L}$ prescription where the spatial zero mode is set to zero on each
time slice,
\begin{align}
a^3\sum_{\mathbf{x}}A_\mu(t,\mathbf{x})=0\,.
\end{align}
Below we will expand the path integral to NLO in $\alpha$. To this end it is
instructive to  first integrate out the scalar fields analytically,
\begin{align}\label{eq:PI_Opexp}
\langle O\rangle = \frac 1{\mathcal{Z}_{\rm L}}\int \mathcal{D}A \, O_{\rm Wick}[\Delta^{-1}]\,{\rm det}(\Delta)^{-\frac 12}\,e^{-S_{{\rm L},A}[A]}\,,
\end{align}
where $O_{\rm Wick}$ represents the observable after the Wick contraction. The
action is symmetric under $A_\mu\to-A_\mu$ and therefore, contributions odd in
$q$ do not contribute to expectation values. To NLO we can therefore set ${\rm
det}(\Delta)=1$.

We rewrite the operator $\Delta$ with the help of the translation operator $\tau
_{\mu} f(x) = f(x+a\hat{\mu})$, as
\begin{equation}
\Delta = {a^{-2}}\Big( 2-e^{iqaA_{\mu}}\tau _{\mu} -\tau _{-\mu} e^{-iqaA_{\mu}}\Big) + m^{2} \,.
\end{equation}
The expansion of $\Delta$ in the electric charge $q$ takes the form,
\begin{align}\label{eq:deltadef}
& \Delta = \Delta _{0} +q\Delta _{1}+q^{2}\Delta _{2} +\ldots ,
\end{align}
where
\begin{gather}
 \Delta _{0} = m^{2}-\frac 1{a^2}\sum _{\mu}\left(\tau_\mu+\tau_{-\mu}-2\right)\,,\qquad
 \Delta _{1} = -\frac{i}{a}\sum _{\mu} \Big( A_{\mu}\tau _{\mu} - \tau _{-\mu}A_{\mu}\Big)  \, , \nonumber \\
\Delta _{2} = \frac{1}{2}\sum _{\mu} \Big( A^{2}_{\mu} \tau _{\mu} + \tau _{-\mu} A^{2}_{\mu}\Big) \label{eq:Delta_i}\,.
\end{gather}
Inserting the kernel $\Delta$ expanded in $q$ into the scalar-QED action,
\begin{align}
S_{\phi} \left[ \phi , A\right] = \frac{a^{4}}{2}\sum _{x}\phi ^{*}(x) \Bigg( \Delta _{0} +q\Delta _{1}+q^{2}\Delta _{2} +\ldots  \Bigg) \phi (x) \, ,
\end{align}
allows us to identify the Feynman rules for the inverse free propagator, the
scalar-photon-vertex  and the scalar tadpole, respectively. In particular, the
scalar propagator in the background field $A_{\mu}$ is then readily given by
\begin{align}\label{eq:propexpansion}
\Delta ^{-1} = \Delta _{0}^{-1} -q\,  \Delta _{0}^{-1}\Delta _{1}\Delta _{0}^{-1} + q^{2}\Delta _{0}^{-1}\Delta _{1}\Delta _{0}^{-1}\Delta _{1}\Delta _{0}^{-1}-q^{2}\Delta _{0}^{-1}\Delta _{2}\Delta _{0}^{-1} +O(q^3)\, .
\end{align}
From this expansion it is a simple exercise to derive the associated Feynman
rules for lattice perturbation theory.
\subsection{Lattice perturbation theory Monte-Carlo strategy}
In order to numerically check the analytic results we numerically  calculate the
finite-size corrections $\Delta \hat{\Pi}_{U}$ for each diagram $U$ in scalar
$\textrm{QED}$ using lattice perturbation theory (LPT). We present below the
analytic expressions for the diagrams (E), (C), (T), (S), and (X) in  lattice
perturbation theory, which are the discrete version of
\cref{eq:intE,eq:intC,eq:intT,eq:intS,eq:intX}.
\begin{align}
(E) : \, \, & \frac{\left(\widehat{ 2\ell-q}\right) _{\mu}\left(\widehat{ 2\ell-q}\right) _{\nu}(\widehat{2\ell+k})^{2}}{\hat{k}^{2}\left( \hat{\ell}^{2}+m^{2}\right) ^{2}\left( \left(\widehat{ k+\ell}\right) ^{2}+m^{2}\right)\left( \left( \widehat{\ell-q}\right) ^{2}+m^{2}\right)  },
  \label{eq:lintE}\\
(C) : \, \, &\frac{-2\left(\widehat{2q- 2\ell-k}\right) _{\mu}\left(\widehat{ q-2\ell }\right) _{\nu}\cos \left( \frac{a}{2}(q-k-2\ell)\right) _{\mu}   }{\hat{k}^{2}\left( \hat{\ell}^{2}+m^{2}\right) \left( \left(\widehat{ k+\ell-q}\right) ^{2}+m^{2}\right) \left( \left( \widehat{\ell-q}\right) ^{2}+m^{2}\right)   },
    \label{eq:lintC}\\
(T) : \, \, &\frac{-\left(\widehat{ 2\ell-q}\right) _{\mu} \left( \widehat{2\ell-q}\right) _{\nu}\sum _{\alpha} \cos \left( \frac{a}{2}2\ell \right) _{\alpha}}{\hat{k}^{2}\left( \hat{\ell}^{2}+m^{2}\right) ^{2}\left( \left( \widehat{ \ell-q}\right) ^{2} +m^{2}\right) },
  \label{eq:lintT}\\
(S) : \, \, & \frac{4\delta _{\mu \nu}\cos \left(\frac{a}{2}(q-k-2\ell)\right) _{\mu}\cos \left(\frac{a}{2}(q-k-2\ell)\right) _{\nu}}{\hat{k}^{2}\left( \hat{\ell}^{2}+m^{2}\right) \left( \left( \widehat{ k+\ell-q}\right) ^{2}+m^{2}\right)  },
  \label{eq:lintS}\\ 
(X) : \, \, & \frac{-\left(\widehat{q- 2\ell}\right) _{\mu} \left(\widehat{q- 2\ell-2k}\right) _{\nu} \left(\widehat{ 2q-2\ell-k}\right) \cdot \left( \widehat{ 2\ell+k}\right)  }{\hat{k}^{2}\left( \hat{\ell}^{2}+m^{2}\right)\left( \left( \widehat{k+\ell}\right) ^{2}+m^{2}\right) \left( \left( \widehat{\ell-q }\right) ^{2} +m^{2} \right) \left( \left( \widehat{k+\ell-q}\right) ^{2}+m^{2}\right)    },
    \label{eq:lintX}
\end{align}
where $\hat{k}_{\mu}=\frac{2}{a}\sin(\frac{ak_{\mu}}{2})$. 

\begin{figure}[t]
  \includegraphics[width=0.45\textwidth]{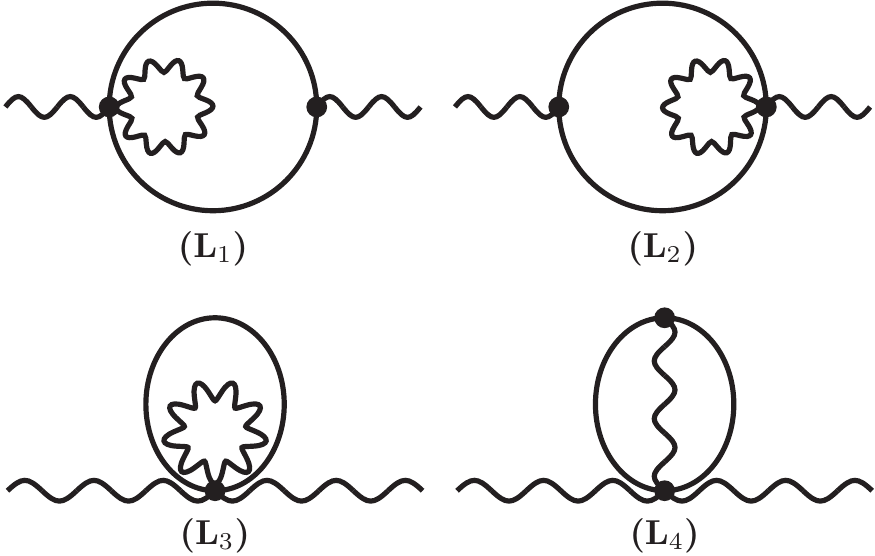}
  \caption{Additional scalar vacuum polarization diagrams specific to lattice perturbation theory.}
  \label{fig:srcsnkTadpole}
\end{figure}

On the lattice, there
is potentially an infinity of new scalar-photon vertices because of the
compactification of the gauge field in~\cref{eq:lcovd}. These vertices are
classically discretisation effects, but at the quantum level they can generate
finite contributions when multiplying power divergences, and ignoring them can
potentially break Ward-Takahashi identities. If one consistently keeps
contributions which do not vanish in the continuum limit, four new diagrams appear
in lattice perturbation theory, represented in~\cref{fig:srcsnkTadpole}. Both
diagrams (L$_{3}$) and (L$_{4}$) are independent of the external momentum and therefore
vanish in the subtracted vacuum polarization function~\cref{eq:PiFVEqn}. 
The
integrand for diagrams (L$_{1}$) and (L$_{2}$) is given by
\begin{align}
  (L) : \, \, &\frac{-\frac{1}{2}a^{2}\left(\widehat{ 2\ell-q}\right) _{\mu} \left( \widehat{2\ell-q}\right) _{\nu}}{\hat{k}^{2}\left( \hat{\ell}^{2}+m^{2}\right) ^{2}\left( \left( \widehat{ \ell-q}\right) ^{2} +m^{2}\right) }\,.\label{eq:lintL}
\end{align}
We integrate these expressions using the VEGAS
algorithm~\citep{1978JCoPh..27..192L}, and more specifically its C++
implementation in the Cuba library~\citep{Hahn:2004fe}. This integration
algorithm builds upon Monte Carlo techniques and creates histograms
approximating the shape of the function which are then used as probability
distributions for importance sampling. This is particularly useful for the
integrals considered here, which are eight-dimensional, and have a complicated
sawtooth-like structure, as we discuss now. In finite volume, the lattice
momentum is discretized and the corresponding sums can be dealt with in VEGAS by
realising that for a function $f(k)$
\begin{equation}\label{eq:vegastrick}
    \sum_{k=0}^{N-1} f(k) = \int _{0}^{N}dk\,f(\lfloor k \rfloor) \, ,
\end{equation}
where $\lfloor k \rfloor$ is the floor operator rounding $k$ down to the nearest
integer. This is extendable to any number of dimensions. As for the
analytic results, we assume an infinite time extent and pions are in infinite
volume, \cf\cref{eq:poissoneqn}, so that only three of the eight integrals are
sums in the finite-volume calculation. The implementation of the calculation is
distributed as a C++ source code under the General Public License v3 in the
supplementary material of this paper, and it features an option to also have the
pions in a finite volume as instructed in the comments. This is particularly
useful when comparing to lattice data, as discussed later.

Each diagram depends on the lattice spacing $a$. 
We introduce a scaling parameter $\sigma$ such that the lattice spacing is
varied according to $a\rightarrow a/\sigma$ and calculate the diagrams for four
different values of $\sigma$, namely $\sigma\in \left\{ 1,1.5,2,3\right\}$, from
which a continuum extrapolation is made by fitting against some polynomial in
$a$. We find, using a pion mass such that $am=0.2$, that the $a$ dependence is
mild. We find the best description of the data in terms of a leading $\mathcal{O}(a^2)$
correction, as expected from the $\mathbb{Z}_2$ symmetry of scalar QED.

Rewriting the sums as in~\cref{eq:vegastrick} yields sawtooth-like behavior
since the integrands of the finite-size effects then are of the form
$f(\ksp)-f(\lfloor \ksp \rfloor)$. The number of discontinuities in this
function is on the order of $(\sigma L/a)^{3}$ (or $(\sigma L/a)^{6}$ if also
the pions are put in finite volume) which means that it can be hard to sample
the integrand efficiently and thus get reliable values and errors from Cuba. The
reliability can be checked by comparing the results from calculations with a
varying number of Monte Carlo evaluation points for a certain $\sigma$. We find
that using $10^{11}$ points gives reliable results for $\sigma <4$. Our result
are summarized at the end of this section.
\subsection{Lattice scalar QED simulations}\label{sec:Numerical}
An alternative avenue which we also explore is to evalute the
lattice-discretized path integral in~\cref{eq:PI_Opexp} by means of a 
Monte Carlo integration for a series of different spatial extents $L$.
This allows for mapping out the volume dependence, thereby checking our
analytical predictions. Instead of numerically solving the momentum sums as in
the previous section one directly samples the path integral
in~\cref{eq:PI_Opexp}. In particular, we compute the vacuum polarization tensor
$\Pi_{\mu\nu}(q)$ as the discrete Fourier transform of the   two-point
function
\begin{align}
C_{\mu\nu}\left(x\right)\equiv\left\langle V_{\mu}\left(x\right)V_{\nu}\left(0\right)\right\rangle\,,
\end{align}
with the lattice conserved vector current
\begin{align}\label{eq:conscurrent}
V_{\mu} (x) =\frac{ i}{ a}[\phi ^{*}(x)U_\mu(x)\phi (x+a\hat{\mu}) -\phi ^{*}(x+a\hat{\mu})U_\mu(x)^{-1}\phi(x)] \, .
\end{align}
After carrying out the Wick contractions we can write the expression for the
vacuum polarization tensor in terms of the propagator in~\cref{eq:propexpansion}
acting on a point source $\delta(x)$,
\begin{align}
    C_{\mu\nu}(x) &= \bigl\langle 2\Re\{[\Delta^{-1}\delta(x)]^{\dagger}U_{\mu}(x)[\tau_\mu\Delta^{-1}\delta(x-a\hat\nu)]U_{\nu}^{\dagger}(0) \nonumber\\
    & \qquad\qquad
    -[\tau_\mu\Delta^{-1}\delta(x)]^{\dagger}U_{\mu}^{\dagger}(x)
    [\Delta^{-1}\delta(x-a\hat\nu)]U_{\nu}^{\dagger}(0)\}\bigr\rangle\,,
\label{eq:vec2pt_withDeltas}\end{align}
where the expectation value represents the functional integration on the gauge
potential $A_{\mu}$. We evaluate this correlation function 
numerically using a setup identical to the one in~\citep{Davoudi:2018qpl}, in
fact the data used here are a side-product of the calculation presented in this
previous work. The covariant Klein-Gordon equation is solved in a stochastic
background field $A_{\mu}$ to form the interacting scalar propagator
$\Delta^{-1}\delta(x)$. Using the expansion~\cref{eq:propexpansion}, this can be
 achieved using the fast Fourier transform algorithm. As a consequence,
this method has a reasonable numerical cost, which is independent from the
chosen scalar mass and has a quasilinear complexity in the number of lattice
points. We refer the reader to~\citep{Davoudi:2018qpl} for more details on the
computational aspects.

\begin{figure}
  \centering
  \includegraphics[width=1\textwidth]{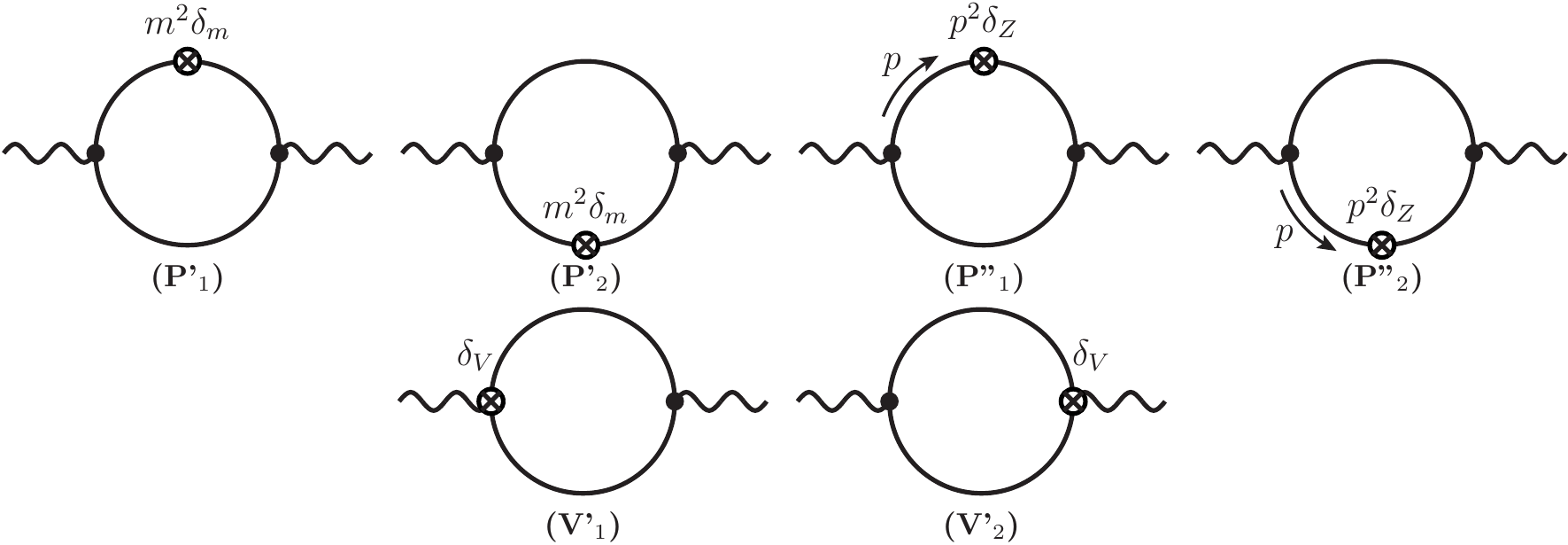} 
  \caption{Counterterm diagrams. The three counterterms $\delta_{m}$, $\delta _{Z}$ and $\delta _{V}$ can be determined by elementary methods.}
  \label{fig:counterdiagrams}
\end{figure}
In principle, the full $\mathcal{O}(\alpha)$ correction to the scalar vacuum
polarization also receives contributions from the diagrams
in~\cref{fig:counterdiagrams}, coming from the 1-loop counter-terms of scalar
QED. We assume that these counterterms are determined through a set of
renormalization conditions in infinite volume, and therefore are independent
of the volume. Because these diagrams do not contain photon propagators, they
clearly do not contribute to~\cref{eq:poissoneqn}. However, the same formula
assumes scalar particles to be in infinite-volume, which is not the case in the
lattice simulation. Although these finite-volume corrections are exponentially
suppressed, they can be greatly enhanced by the ultraviolet-divergent values of
the counter-terms. We therefore included these diagrams to ensure that
exponential finite-volume corrections are negligible for reasonably large values
of $mL$ (the typical threshold for lattice QCD simulations is $M_{\pi}L>4$). The
details of the renormalisation prescription used here are given
in~\cref{app:renormalisation}. The cost of computing the extra counter-term
diagrams is negligible, since they do not depend on the gauge field.
\subsection{Numerical results}
\begin{figure}
  \includegraphics[width=\textwidth]{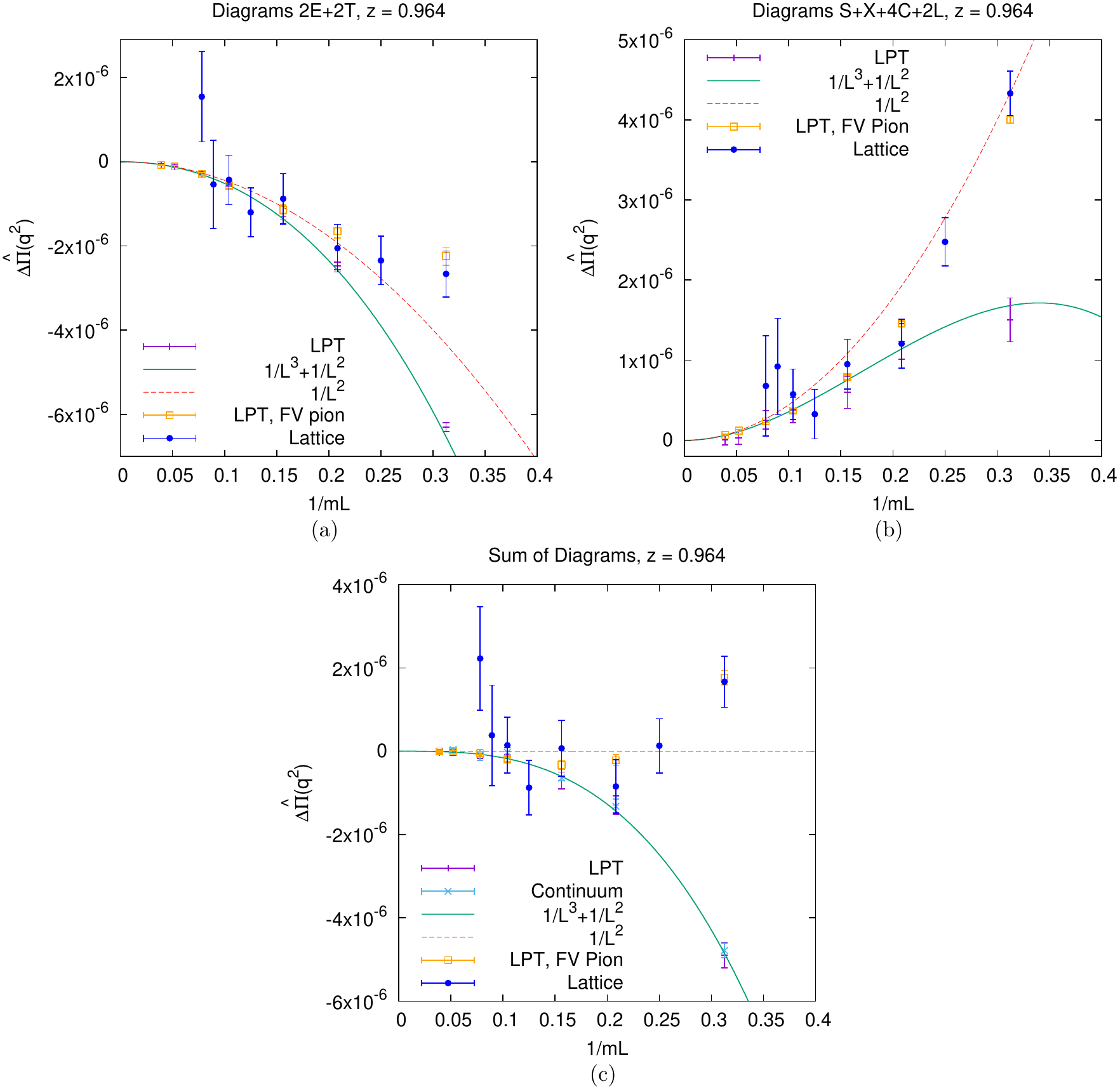}
  \caption{A comparison between the analytic results, LPT and lattice data for (a) $2E+2T$, (b) $S+X+4C$ and (c) $2E+4C+2T+S+X+2L$.}
  \label{fig:numerical}
\end{figure}

In~\cref{fig:numerical} we compare the analytic results to LPT and lattice data.
We use $am=0.2$ and $aq_{0}=8\pi/128$, i.e. $z = q_{0}^{2}/m^{2} \approx 0.964$. The
red dashed line is the $1/L^{2}$ term and the green solid line is the full
expression of the form $1/L^{2}+1/L^{3}$ in the corresponding analytic
expression in~\cref{eq:hvpFV_diagrams}. The purple points are the infinite
volume pion LPT points for a finite $a$, and the crossed blue points are the
continuum extrapolated values. The orange box shaped points are finite volume
pion LPT data. From the infinite volume pion LPT data we clearly see that the
full analytic form is much better than when including only the $1/L^2$ term and
the agreement is excellent up to $1/mL<0.3$ for all diagrams. Other values of
$z$ yield a similar level of agreement. We see that the lattice data starts to
deviate from the analytic curve after $1/mL> 0.2$, but the finite volume pion
calculation reproduces precisely this behavior. We thus attribute the
discrepancies to the exponential finite-size effects that are neglected
in~\cref{eq:poissoneqn} for the analytic calculation as well as in the infinite
volume pion LPT Monte-Carlo. Moreover, for $mL\simeq 4$ we found that the
difference between the infinite-volume pion and finite-volume pion data is on
the order of $10^{-6}$, an order of magnitude smaller than the naive suppression
from a factor of $\alpha$ between the LO and the NLO HVP, viz.
$\hat{\Pi}^{(1)}\sim\alpha\hat{\Pi}^{(0)}\sim\alpha 10^{-3}\sim 10^{-5}$.

\section{Conclusions}
We have performed a 2-loop calculation of the $\mathcal{O}(\alpha)$ corrections
to the hadronic vacuum polarization in scalar QED. We presented the infinite
volume results in terms of 2-loop master integrals from which we obtained an
analytic expression for the finite volume correction to the HVP at this order.
We found that even though each of the individual diagrams contributes as
$1/L^2$, these terms all cancel when combined. We argued that this cancellation
is expected on physical grounds for neutral currents and show that it does not
occur for charged currents. We also argued that this cancellation is universal,
i.e., it occurs regardless of the effective theory used to derive this result.

All our results were tested numerically using two different approaches - direct
integration of lattice perturbation theory integrals using VEGAS and lattice
scalar U(1) gauge theory with stochastically generated photon fields. We find
good agreement between analytic results and results from both numerical
approaches. While absent from our analytical expressions, exponentially
suppressed finite volume effects are visible in our results from the lattice
simulation

Finally, an important consequence of this work is that  for the foreseeable
future finite-volume effects on the QED corrections to the hadronic vacuum
polarization are likely to be negligible in  lattice simulations. For instance,
we expect this to hold even if lattice computations aimed at matching
experimental projections of a four-fold reduction in the error on $a_\mu$ by
Fermilab~\citep{Logashenko:2015xab} and J-PARC~\citep{Otani:2015lra} down to
0.14ppm. This assumes a typical lattice simulation where the pion mass times the
spatial extent is larger than four, for which this work estimates the
finite-size effects to be at the level of only a few percent of the
$\bigo(\alpha)$ correction to the HVP function $\hat{\Pi}(q^2)$. Unless these
effects come with an unnaturally large coefficient in the full theory, they
should be negligible compared to the per-mil accuracy required on the HVP
contribution to $a_{\mu}$. Of course, large coefficients cannot be excluded
considering how critical it is to properly estimate the theoretical uncertainty
on $a_{\mu}$, particularly in the perspective of confirming or excluding the
current discrepancy between experiment and theory on this quantity. The results
of this work together with simulations of full lattice QCD+QED even with a
limited number of volumes, should allow to constrain the size of these
effects. 

  \begin{acknowledgments}
    A.P. would like to thank the High Energy Physics Group of Lund University
    for its warm welcome, important parts of this work were initiated during
    A.P.'s visit in Lund. Lattice computations presented in this work have been
    performed on DiRAC equipment which is part of the UK National
    E-Infrastructure, and on the IRIDIS High Performance Computing Facility at
    the University of Southampton. T.J. and A.P. are supported in part by UK
    STFC grants ST/L000458/1 and ST/P000630/1. A.P. also received funding from
    the European Research Council (ERC) under the European Union's Horizon 2020
    research and innovation programme under grant agreement No 757646. J.B. and
    N.H.T. are supported in part by the Swedish Research Council grants contract
    numbers 2015-04089 and 2016-05996, and by the European Research Council
    (ERC) under the European Union's Horizon 2020 research and innovation
    programme under grant agreement No 668679. J.H. was supported by the EPSRC
    Centre for Doctoral Training in Next Generation Computational Modelling
    grant EP/L015382/1. A.J. received funding from STFC consolidated grant
    ST/P000711/1 and from the European Research Council under the European
    Union's Seventh Framework Program (FP7/2007- 2013) / ERC Grant agreement
    279757. 
  \end{acknowledgments}
\appendix
\section{In continuous infinite volume}\label{sec:infv}
In this section the HVP is considered in continuous infinite volume Minkowski space. We calculate $\Pi ^{(0)}\left( q^{2}\right) $ and $\Pi ^{(1)}\left( q^{2}\right) $ in $\overline{MS}$ and numerically compare their respective sizes. The corresponding calculation in QED can be found in~\citep{Kallen:1955fb,Barbieri:1973mt}. 

In Minkowski space the scalar QED Lagrangian is
\begin{align}
\mathcal{L} = \left(\partial_\mu\phi^*+i e A_\mu\phi^*\right)
\left(\partial^\mu\phi-i e A^\mu\phi\right)-m^2\phi^*\phi-\frac{1}{4}F_{\mu\nu}F^{\mu\nu} \, .
\label{eq:LagrangianMinkEqn}\end{align}
The relevant counterterms for the parameters above, are defined in the counterterm Lagrangian
\begin{align}
\mathcal{L}_{CT} = -ie\delta _{\phi ^{2}A}\left(\partial_\mu\phi^{*}\phi-\phi^{*}\partial _{\mu}\phi\right) A^{\mu}+e^{2}\delta _{\phi ^{2}A^{2}} A_{\mu}\phi ^{*}A^{\mu}\phi -m^2\delta _{m}\phi^*\phi-\frac{1}{4}\delta _{F}F_{\mu\nu}F^{\mu\nu} \, .
\end{align}
In $d=4-2\varepsilon$ dimensions these are given by
\begin{align}
\delta _{F} = -\frac{1}{48\pi^{2}}\frac{1}{\epsilon},  \; \; \;  \delta _{\phi} =\delta _{\phi ^{2}A} = \delta _{\phi ^{2}A^{2}} =  \frac{1}{8\pi^{2}}\frac{1}{\epsilon}, \; \; \; \delta _{m} = -\frac{1}{16\pi^{2}}\frac{1}{\epsilon}.
\end{align}  
Note that diagrams (A) and (T) identically vanish in dimensional regularization, so that we are left with diagrams (F), (G), (D), (B), (E), (C), (S) and (X). The two HVP contributions can thus be written (cf. the FV case in~\cref{eq:combinatoricPiFV})
\begin{align}\label{eq:combinatoricPi}
\Pi ^{(0)}\left( q^{2}\right) =\, & (F)+(G), \nonumber \\
\Pi ^{(1)}\left( q^{2}\right)  =\, & (B) + 2\cdot (E) + 4\cdot (C) + (S) + (X)+(D).
\end{align}
Using the tensor structure of $\Pi ^{\mu \nu} \left( q^{2}\right) $ and the Ward identity it is easy to see that the disconnected part is given by the squared LO contribution, $(D) = \left( \Pi ^{(0)}\left( q^{2}\right) \right) ^{2}$. The diagrams are given in~\cref{sec:MinkDiagApp}.

Using Lorentz invariance identities and integration by parts, the 2-loop integrals can be rewritten in a basis of master integrals. The program \textsc{Reduze2}~\citep{vonManteuffel:2012np} employs a Laporta algorithm in order to do this, and allows the user to define such a basis. The master integrals used here are the $\overline{MS}$ subtracted parts of
\begin{align}\label{eq:MasterIntsEqn}
&A(m^2) =\,\frac{1}{i}\int\frac{d^d\ell}{(2\pi)^d}\frac{1}{\ell^2-m^2},
\nonumber\\
&B(m^2,q^2) =\,\frac{1}{i}\int\frac{d^d\ell}{(2\pi)^d}\frac{1}{(\ell^2-m^2)\left((\ell-q)^2-m^2)\right)},
\nonumber\\
 &S(m^2,q^2)=\,\frac{1}{i^2}\int\frac{d^d\ell}{(2\pi)^d}\frac{d^dk}{(2\pi)^d}
\frac{1}{k^2(\ell^2-m^2)\left((k+\ell-q)^2-m^2)\right)},
\nonumber\\
&T(m^2,q^2)=\,\frac{1}{i^2}\int\frac{d^d\ell}{(2\pi)^d}\frac{d^dk}{(2\pi)^d}
\frac{1}{k^2(\ell^2-m^2)^2\left((k+\ell-q)^2-m^2)\right)},
\nonumber\\
&V(m^2,q^2)=\,\frac{1}{i^2}\int\frac{d^d\ell}{(2\pi)^d}\frac{d^dk}{(2\pi)^d}
\frac{1}{k^2(\ell^2-m^2)^2\left((k+\ell)^2-m^2\right)\left((\ell-q)^2-m^2)\right)},
\nonumber\\
&M(m^2,q^2)=\,\frac{1}{i^2}\int\frac{d^d\ell}{(2\pi)^d}\frac{d^dk}{(2\pi)^d}
\frac{1}{k^2(\ell^2-m^2)\left((k+\ell)^2-m^2\right)\left((\ell-q)^2-m^2)\right)
\left((k+\ell-q)^2-m^2\right)}.
\end{align}
All but integral $M$ are divergent and thus require expansion in $\varepsilon$ in order to isolate the divergent parts from the finite ones, something which can be done in both Euclidean and Minkowski space. For a Euclidean spacetime the analytic results can be found in~\citep{Martin:2003qz}. However, working in Minkowski space, the corresponding expressions are here given in~\cref{sec:MasterIntegralsApp}.

\begin{figure}[htbp]
\includegraphics[width=\textwidth]{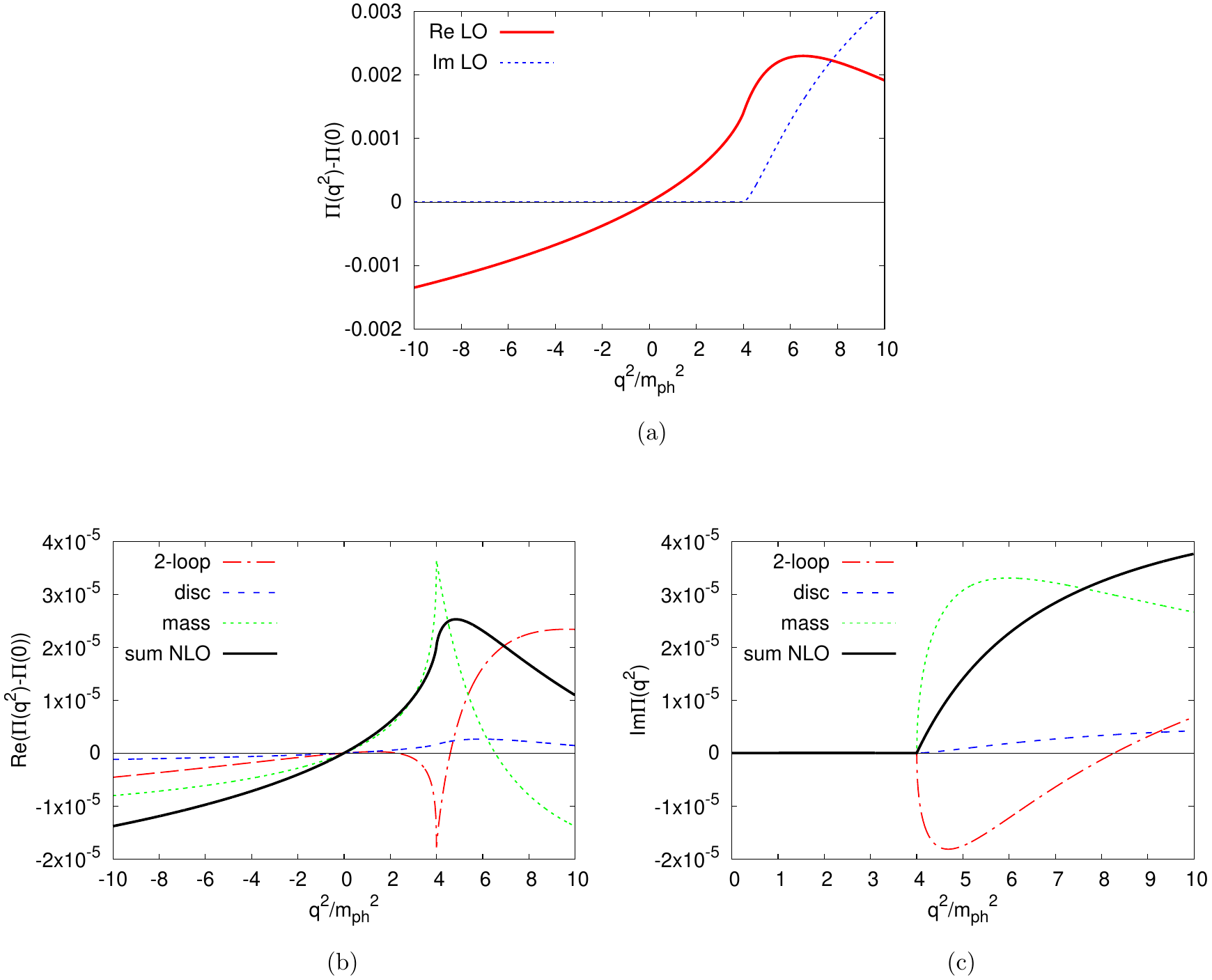}
\caption{The various contributions to the scalar vacuum polarization in an infinite volume with Minkowski signature: (a)~LO, (b)~Real part of NLO, (c)~Imaginary part of NLO.}
\label{fig:HVPIVfigplots}
\end{figure}

Note that in $\overline{MS}$ the threshold shift can, and does, induce a sign change of the imaginary part of $\Pi ^{(1)}\left( q^{2}\right) $ at some $q^{2}$. However, this does not occur for an on-shell scheme or with the physical mass. The physical mass $m_{ph}^{2}$ is related to $m^{2}$ through
\begin{align}
m_{ph}^{2} = m^{2}+\frac{\alpha }{4\pi } m^{2}\left( 7-3\log \frac{m^{2}}{\mu ^{2}}\right) \equiv m^{2}+\delta m^{2} .
\end{align}
The HVP can therefore also be expanded around this mass, 
\begin{align}
\Pi \left( q^{2}\right) = \, & \left. \Pi ^{(0)}\left( q^{2}\right)  \right| _{m^{2}=m_{ph}^{2}} + \delta m^{2} \frac{\partial }{\partial m^{2}} \left. \Pi ^{(0)}\left( q^{2}\right)  \right| _{m^{2}=m_{ph}^{2}} + \left. \Pi ^{(1)}\left( q^{2}\right)  \right| _{m^{2}=m_{ph}^{2}} + \ldots \nonumber \\ \equiv \, & \Pi ^{\text{1-loop}} + \Pi ^{\delta m^{2}} + \Pi ^{\text{disc}}+ \Pi ^{\text{2-loop}}+\ldots \, ,
\end{align}
where in the last step the disconnected part was separated from the NLO contribution. To simplify the expressions, let us further define
\begin{align}
\sigma^2 =\,& 1-\frac{4m_{ph}^2}{q^2}.
\end{align}
The HVP contributions at LO and NLO are thus
\begin{align}
\Pi^{\text{1-loop}} =\,&\frac{4}{3}\overline{A}(m_{ph}^2)+\frac{1}{3}\sigma^2
\overline{B}(m_{ph}^2,q^2)+\frac{1}{16\pi^2}\left(\frac{2}{9}-\frac{4m_{ph}^2}{3q^2}\right)
\, , \nonumber\\
\Pi^{\delta m^2} =\,&-\delta m^2 \frac{2}{q^2}\left(\frac{1}{m_{ph}^2}\overline{A}(m_{ph}^2)-\overline{B}(m_{ph}^2,q^2)-\frac{1}{16\pi^2}\right)
\, , \nonumber\\
\Pi^{\text{disc}} =\,&\left(\Pi^{\text{1-loop}}\right)^2
\, , \nonumber\\
\Pi^\text{2-loop} = \,& \frac{1}{(16\pi^2)^2}\left(\frac{10}{3}-\frac{8m_{ph}^2}{q^2}\right)
+\frac{\overline{A}(m_{ph}^2)}{16\pi^2}\frac{22}{3q^2}+\frac{10}{3m_{ph}^2q^2}\overline{A}(m_{ph}^2)^2
\nonumber\\&
+\left(\frac{8}{3m_{ph}^2}-\frac{26}{3q^2}\right)\overline{A}(m_{ph}^2)\overline{B}(m_{ph}^2,q^2)
+\frac{8\sigma^2}{48\pi^2}\overline{B}(m_{ph}^2,q^2)
\nonumber\\&
-\frac{8}{3}\left(\frac{1}{q^2}\overline{S}(m_{ph}^2,q^2)+\sigma^2\overline{T}(m_{ph}^2,q^2)-m_{ph}^2\sigma^2\overline{V}(m_{ph}^2,q^2)\right)
\nonumber\\&
+\left(-\frac{4}{3}+\frac{8m_{ph}^2}{3q^2}\right)\overline{B}(m_{ph}^2,q^2)^2
-\frac{2\sigma^2}{3}\left(q^2-\frac{2m_{ph}^2}{q^2}\right)\overline{M}(m_{ph}^2,q^2)
\, , 
\end{align}
where the quantities with bars are the finite parts of the integrals in~\cref{eq:MasterIntsEqn}. These contributions as well as the corresponding subtracted quantities are plotted in~\cref{fig:HVPIVfigplots} for $m_{ph} = 139.5$ MeV, $\mu = 500$ MeV and $e=0.303$. As can be seen, the NLO parts are roughly two orders of magnitude smaller than LO, this is due to the additional power of $\alpha \sim 10^{-2} $. Moreover, it can be noted that $\Pi ^{\text{disc}}$ on average is significantly smaller than the other parts, and that $\Pi^{\text{disc}}$ and $\Pi^\text{2-loop}$ combine to give the proper non-singular threshold behavior. 

\subsection{Diagrams in Minkowski space}\label{sec:MinkDiagApp}
Using the Feynman rules for the Lagrangian in~\cref{eq:LagrangianMinkEqn}, the diagrams are 
\begin{align}
(F) =\, & \int \frac{d^{d}\ell}{\left( 2\pi \right) ^{d}} \frac{-2ig^{\mu \nu }}{\ell^{2}-m^{2}},
 \nonumber \\
(G) =\, & \int \frac{d^{d}\ell}{\left( 2\pi \right) ^{d}}\frac{i\left( 2\ell-q\right)^{\mu} \left( 2\ell-q\right) ^{\nu}  }{\left( \ell^{2}-m^{2}\right) \left( \left( \ell-q\right) ^{2}-m^{2}\right)  },
\nonumber \\
(A) = \, &\int \frac{d^{d}\ell}{\left( 2\pi \right) ^{d}}\frac{d^{d}k}{\left( 2\pi \right) ^{d}} \frac{2idg^{\mu \nu}}{k^{2}\left( \ell^{2} - m^{2}\right) ^{2}}, \nonumber \\
(B) =\, &\int \frac{d^{d}\ell}{\left( 2\pi \right) ^{d}}\frac{d^{d}k}{\left( 2\pi \right) ^{d}} \frac{-2i g^{\mu \nu}(2\ell+k)^{2}}{k^{2}\left( \ell^{2}-m^{2}\right) ^{2}\left( \left( k+\ell\right) ^{2} - m^{2} \right)  },
\nonumber \\
(E) =\, & \int \frac{d^{d}\ell}{\left( 2\pi \right) ^{d}}\frac{d^{d}k}{\left( 2\pi \right) ^{d}} \frac{i\left( 2\ell-q\right) ^{\mu}\left( 2\ell-q\right) ^{\nu}(2\ell+k)^{2}}{k^{2}\left( \ell^{2}-m^{2}\right) ^{2}\left( \left( k+\ell\right) ^{2}-m^{2}\right)\left( \left( \ell-q\right) ^{2}-m^{2}\right)  },
\nonumber \\
(C) =\, &\int \frac{d^{d}\ell}{\left( 2\pi \right) ^{d}}\frac{d^{d}k}{\left( 2\pi \right) ^{d}}  \frac{-2i\left( 2\ell+k\right) ^{\mu}\left( 2\ell-q\right) ^{\nu}  }{k^{2}\left( \ell^{2}-m^{2}\right) \left( \left( k+\ell\right) ^{2}-m^{2}\right) \left( \left( \ell-q\right) ^{2}-m^{2}\right)   },
 \nonumber  \\
(T) =\, & \int \frac{d^{d}\ell}{\left( 2\pi \right) ^{d}}\frac{d^{d}k}{\left( 2\pi \right) ^{d}} \frac{-id\left( 2\ell-q\right) ^{\mu} \left( 2\ell-q\right) ^{\nu}}{k^{2}\left( \ell^{2}-m^{2}\right) ^{2}\left( \left( \ell-q\right) ^{2} -m^{2}\right) },
\nonumber \\
(S) =\, &  \int \frac{d^{d}\ell}{\left( 2\pi \right) ^{d}}\frac{d^{d}k}{\left( 2\pi \right) ^{d}} \frac{4ig^{\mu \nu}}{k^{2}\left( \ell^{2}-m^{2}\right) \left( \left( k+\ell-q\right) ^{2}-m^{2}\right)  },
 \nonumber \\
(X) =\, & \int \frac{d^{d}\ell}{\left( 2\pi \right) ^{d}}\frac{d^{d}k}{\left( 2\pi \right) ^{d}} \frac{i\left( 2\ell-q\right) ^{\mu} \left( 2\ell+2k-q\right) ^{\nu} \left( 2\ell+k-2q\right) \cdot \left( 2\ell+k\right)  }{k^{2}\left( \ell^{2}-m^{2}\right)\left( \left( k+\ell\right) ^{2}-m^{2}\right) \left( \left( \ell-q \right) ^{2} -m^{2} \right) \left( \left( k+\ell-q\right) ^{2}-m^{2}\right)    }.
\nonumber  
\end{align} 

\subsection{Master integrals}\label{sec:MasterIntegralsApp}
Below, each Minkowski-space master integral in~\cref{eq:MasterIntsEqn} has been separated into a finite and an infinite part, the finite one denoted by a bar. The analytic expressions for these finite integrals are given in~\citep{Martin:2003qz}, and are in terms of Riemann zeta functions as well as the polylogarithm functions $\textrm{Li}_{2}$ and $\textrm{Li}_{3}$,
\begin{align}
A(m^2) =\,&\frac{m^{2}}{16\pi^{2} \varepsilon}+\overline{A}\left( m^{2}\right) ,
\nonumber\\
B(m^2,q^2) =\,&\frac{1}{16 \pi ^{2}\varepsilon}+\overline{B}\left( m^{2},q^{2}\right) ,
\nonumber\\
S(m^2,q^2)=\,& -\frac{3m^{2}}{512\pi ^{4}\varepsilon ^{2}}+\frac{-q^{2}+6m^{2}}{512\pi^{4}\varepsilon}+\frac{3}{16\pi^{2}\varepsilon}A\left( m^{2}\right) + \overline{S}\left( m^{2},q^{2}\right) ,
\nonumber\\
T(m^2,q^2)=\,& -\frac{1}{512\pi ^{4}\varepsilon^{2}}+\frac{1}{512\pi ^{4}\varepsilon}+\frac{1}{16\pi ^{2}\epsilon}\frac{\left( 1-\varepsilon\right)}{m^{2}} A\left( m^{2}\right) +\overline{T}\left( m^{2},q^{2}\right) ,
\nonumber\\
V(m^2,q^2)=\, & \frac{1}{16\pi ^{2}\varepsilon}\Bigg[ \left( \frac{d-3}{4m^{2}-q^{2}}\right) B\left( m^{2}, q^{2}\right) + \frac{2-d}{\left( 4m^{2}-q^{2}\right) q^{2} }A\left( m^{2}\right) 
\nonumber \\ & +\frac{\left( d-2 \right) \left( 2m^{2}-q^{2}\right) }{2\left( 4m^{2}-q^{2}\right) m^{2}q^{2}}A\left( m^{2}\right) \Bigg] +\overline{V}\left( m^{2},q^{2}\right) ,
\nonumber\\ 
M(m^2,q^2)=\,& \overline{M}\left( m^{2},q^{2}\right) .
\label{eq:masterints}\end{align}

\section{The scalar loop integrals $\Omega _{\alpha,\beta}$}\label{sec:OmegaApp}
Consider the dimensionless function $\Omega _{\alpha , \beta}$ given by
\begin{equation}
  \Omega_{\alpha,\beta}(z)=\frac{1}{2\pi^2}\int_0^{\infty}d x\,
  x^2\omega_{\alpha,\beta}(x,z) \, ,
\end{equation}
where
\begin{equation}
  \omega_{\alpha,\beta}(x,z)=
  \frac{1}{(x^2+1)^{\frac{\alpha}{2}}[z+4(x^2+1)]^{\beta}}\,.
  \label{eq:Omdef}
\end{equation}
This function converges if and only if $\alpha+2\beta>3$. The relations between
$m$, the external momentum $q^{2}$, the integration variable $x$ and the
variable $z$ are $z=q^{2}/m^2$ and $x = \sqrt{\ellsp ^{2}} /m = \ell/m$. It is
also possible to write $\Omega _{\alpha, \beta}$ explicitly in terms of
hypergeometric functions as
\begin{align}
  \Omega_{\alpha,\beta}(z)&=
  \frac{8\sqrt{\pi}(\beta-2)(z+4)^{\frac{3}{2}-\beta}}{z\Gamma\left(\frac{5}{2}-\beta\right)\Gamma(\beta)}\,_2F_1\left(-\frac{1}{2},\frac{\alpha}{2};\frac{5}{2}-\beta;\frac{z}{4}+1\right)
  \notag\\
  &\quad
  +\frac{\sqrt{\pi}(z+4)^{\frac{3}{2}-\beta}[z-4(\alpha+2\beta-4)-\alpha z]}{z\Gamma\left(\frac{5}{2}-\beta\right)\Gamma(\beta)}
  \,_2F_1\left(\frac{1}{2},\frac{\alpha}{2};\frac{5}{2}-\beta;\frac{z}{4}+1\right)
  \notag\\
  &\quad
  -\frac{4^{2-\beta }\Gamma\left(\frac{\alpha}{2}+\beta-\frac{3}{2}\right)}{\Gamma\left(\frac{\alpha}{2}\right)\Gamma\left(\beta-\frac{1}{2}\right)}\,_2F_1\left(\beta,\frac{\alpha-3}{2}+\beta;\beta-\frac{1}{2};\frac{z}{4}+1\right)\, ,
  \label{eq:omhyper}
\end{align}
where $\, _2F_1$ is the hypergeometric function defined by
\begin{equation}
  \, _2F_1(a,b;c;z)=\frac{\Gamma(c)}{\Gamma(b)\Gamma(c-b)}
  \int_0^{\infty}d x\,x^{-b+c-1}(x+1)^{a-c}(x-z+1)^{-a}\,.
\end{equation}
However, the form~\cref{eq:omhyper} is complex and might not be the most useful
in practice. The $\Omega_{\alpha,\beta}$ functions are actually related to each
other and can be expressed as combinations of a smaller set of functions. One
starts by noticing the relations
\begin{align}
  \frac{\partial}{\partial z}\omega_{\alpha\beta}(x,z)&=
  -\beta\omega_{\alpha,\beta+1}(x,z)\label{eq:domdz}\,,\\
  \frac{\partial}{\partial x}\omega_{\alpha\beta}(x,z)&= -\alpha
  x\omega_{\alpha+2,\beta}(x,z) -8\beta
  x\omega_{\alpha,\beta+1}(x,z)\label{eq:domdx}\,.
\end{align}
The identity~\cref{eq:domdz} directly implies that
\begin{equation}
  \Omega_{\alpha,\beta+1}(z)=-\frac{1}{\beta}\frac{\partial}{\partial
  z}\Omega_{\alpha\beta}(z)\,,
  \label{eq:ombrec}
\end{equation}
i.e. the index $\beta$ is decreased by taking derivatives in $z$. One can
further note that
\begin{align}
\Omega _{\alpha , \beta}(z) = z\, \Omega _{\alpha , \beta +1}(z)+4\, \Omega _{\alpha -2, \beta +1}(z) \, ,
\end{align}
which is easily proven by multiplying and dividing the integrand a factor
$z+4(x^{2}+1)$. For the case $\alpha +2\beta >3$ it is possible to find a
recursion relation by using~\cref{eq:domdx}, this by partially integrating the
definition of $\Omega _{\alpha,\beta}$:
\begin{align*}
\Omega _{\alpha,\beta} (z) = & \frac{1}{2\pi^2}\int_0^{\infty}d x\,
  x^2\omega_{\alpha\beta}(x,z) = \Bigg[ \frac{x^{3}}{3}\omega _{\alpha , \beta}(x,z)\Bigg] _{0}^{\infty} - \frac{1}{3}\int _{0}^{\infty}d x \, x^{3}  \frac{\partial}{\partial x}\omega_{\alpha\beta}(x,z) \nonumber 
  \\
  =&  \frac{1}{3}\int _{0}^{\infty}d x \, x^{4} \Big\{ \alpha \,  \omega_{\alpha+2\beta}(x,z)+8\beta \,  \omega_{\alpha\beta+1}(x,z)\Big\} \, ,
\end{align*}
where in the last step the convergence requirement $\alpha +2\beta >3$ as well
as~\cref{eq:domdx} were used. By writing $x^{4} = x^{2}\Big( x^{2}+1\Big)-x^{2}
$ one then finds the relation
\begin{align}
\Omega _{\alpha , \beta}(z) = \frac{1}{3}\Big\{ \alpha \, \Omega_{\alpha,\beta}(x)- \alpha \, \Omega_{\alpha+2,\beta}(x) +8\beta \, \Omega_{\alpha-2,\beta +1}(x)-8\beta \, \Omega_{\alpha,\beta +1}(x)\Big\} \,  ,
\end{align}
or, for $\alpha >0$,
\begin{align}
\Omega _{\alpha +2 , \beta}(z) =  \frac{\alpha -3}{\alpha} \Omega_{\alpha,\beta}(x) +\frac{8\beta}{\alpha} \Omega_{\alpha-2,\beta +1}(x)- \frac{8\beta}{\alpha} \Omega_{\alpha,\beta +1}(x) \,  .
\end{align}

Inspired by the recursion relation in~\cref{eq:ombrec}, define for $\alpha>1$
the functions
\begin{equation}
  \Omega_{\alpha}(z)=\Omega_{\alpha,1}(z)\,.
\end{equation}
Now, using~\cref{eq:ombrec} for $\alpha>1$ and an integer $\beta\geq 1$
\begin{equation}
  \Omega_{\alpha , \beta}(z)=\frac{1}{(\beta-1)!}
  \left(-\frac{\partial}{\partial z}\right)^{\beta-1}
  \Omega_{\alpha}(z)\,.\label{omaboma}
\end{equation}
The second identity~\cref{eq:domdx}, combined with an integration by parts of
\cref{eq:Omdef,eq:ombrec} leads to
\begin{equation}
  \Omega_{\alpha+4}(z)=
  \left(1-\frac{3}{\alpha+2}\right)\Omega_{\alpha+2}(z)
  +\frac{8}{\alpha+2}\Omega'_{\alpha+2}(z)
  -\frac{8}{\alpha+2}\Omega'_{\alpha}(z)\,,
\end{equation}
where we used the prime notation for derivatives in $z$. Using this last
relation and~\cref{omaboma}, it is clear that for any positive integer couple
$(\alpha,\beta)$ such that $\alpha+2\beta>3$, $\Omega_{\alpha\beta}(z)$ is a
linear combination of the six following functions and their derivatives
\begin{align}
  \Omega_{2,1}(z)&=\frac{1}{8\pi z}\left(\sqrt{4+z}-2\right)\,,\\
  \Omega_{3,1}(z)&=\frac{1}{4\pi^2z}\left\{\sqrt{1+\frac{4}{z}}\log\left[\frac{1}{2}\left(z+\sqrt{z(z+4)}+2\right)\right]-2\right\}\,,\\
  \Omega_{4,1}(z)&=\frac{1}{8\pi z^2}\left(z-4\sqrt{z+4}+8\right)\,,\\
  \Omega_{5,1}(z)&=\frac{1}{6\pi^2z^{\frac52}}\left\{z^{\frac32}+12\sqrt{z}-6\sqrt{z+4}\log\left[\frac{1}{2}\left(z+\sqrt{z(z+4)}+2\right)\right]\right\}\,,\\
  \Omega_{0,2}(z)&=\frac{1}{64\pi\sqrt{z+4}}\,,\\
  \Omega_{1,2}(z)&=\frac{1}{16\pi^2z^2(z+4)}\left\{z(z+4)-2\sqrt{z(z+4)}\log\left[\frac{1}{2}\left(z+\sqrt{z(z+4)}+2\right)\right]\right\}\,.
\end{align}

\section{Lattice scalar QED renormalisation scheme}\label{app:renormalisation}
We start by rewriting the Lagrangian of scalar QED in term of the renormalized
fields and parameters defined by $\phi_0 = \sqrt{Z_\phi}\phi$, $A^\mu_0 =
\sqrt{Z_A}A^\mu_0$, $m = Z_m m$, $e_{0} = Z_e e$, where a subscript 0 denotes a
bare quantity. The counterterm part of the lattice Lagrangian is given by
\begin{align}
    \mathcal L_{ct} &= \underbrace{(Z_\phi - 1)}_{\delta_Z} |\delta_\mu \phi|^2 + \underbrace{(Z_m Z_\phi - 1)}_{\delta_m} m^2 |\phi|^2
    + iq\underbrace{(Z_q Z_\phi \sqrt{Z_A} - 1)}_{\delta_V} A_\mu [\phi^* \delta_\mu \phi - (\delta_\mu\phi)^*\phi]\nonumber\\
    &\quad+q^2(Z_q^2Z_AZ_\phi - 1)|\phi|^2{\textstyle\sum_{\mu}}A_\mu^2
     + \frac{1}{4}(Z_A - 1)\,{\textstyle\sum_{\mu\nu}}F_{\mu\nu}^2
     + \frac{1}{2}(Z_A - 1)\,{\textstyle\sum_{\mu}}(\delta_{\mu}A_{\mu})^2
\end{align}
At the order $\mathcal{O}(q^2)$ relevant here, the electric charge $q$ does not
renormalize, \ie $Z_A=1$. The discretized action is gauge invariant and, as it
is well known in the continuum, the theory can be renormalized by removing
divergences in the self-energy function and by using $\delta_V=\delta_Z$ as
imposed by the Ward-Takahashi identities. By denoting $\Sigma(p)$ the
self-energy function at momentum $p$, we choose the following renormalization
prescription
\begin{equation}
  \Sigma(0)=0\qquad\text{and}\qquad\Sigma(q_T)=0\,,
\end{equation}
with $q_T=(\frac{2\pi}{T},\mathbf{0})$ where $T$ is the time extent of the
lattice. This prescription allows to compute the wave function renormalization at
finite time extent. In all the finite-time numerical results presented in this
paper we used $T=128a$. For $T\to\infty$, this prescription gives back the more
traditional conditions, where one assumes that the self-energy and its
derivative vanishes at $p^2=0$. For $T=128a$ and $am=0.2$ we found
\begin{equation}
  a^2 m^2\delta_m=-0.466819(2)q^2\qquad\text{and}\qquad\delta_Z=0.146054(4)q^2\,.
\end{equation}

\section{Explicit forms of energy-integrated diagrams}\label{sec:diagramapp}
The subtracted functions $\hat{\rho}_{U}$ can be written in the form
\begin{align}\label{eq:tildepiapp}
& \hat{\rho}_{U} (\mathbf{k},\ellsp , q_{0}) = C^{U}\sum _{i = 0}^{1}\sum _{j = 0}^{5} A_{i}^{U}a_{ij}^{U}\left| \mathbf{k}\right| ^{j-1}  \, , 
\end{align}
where $C^{U}$, $A^{U}_{i}$ and $a_{ij}^{U}$ are functions of $\mathbf{k}$,
$\ellsp$ and $q_{0}$. The above factorization is chosen such that the dependence
on $\mathbf{k}$ in these functions is different from pure powers of $\left|
\mathbf{k}\right|$. This means that they can depend on $\mathbf{k}$ in
denominators through the energy $\omega _{p}=\sqrt{\mathbf{p}^2+m^2}$ (which
often shows up in denominators) as well as the combination $\vdku $ for the
velocity $\vel=\frac{{\ellsp}}{\omega _{\ell}}$ and the unit vector
$\hat{\ksp}=\frac{\ksp}{|\ksp|}$. This separation is useful since, for a given
$j$, a large volume expansion of $C^{U}A_{i}^{U}a_{ij}^{U}$, which multiplies
$\left| \mathbf{k}\right| ^{j-1}$, has leading power behavior of order $\left|
\mathbf{k}\right| ^{j-1}$. It is therefore only the $j=0$ term in the sum over
$j$ which can give a contribution to $b_{1}^{U}$ and thus a $1/L^{2}$
finite-size correction, where the coefficients $b_{1}^{U}$ and $b_{0}^{U}$ are
defined through
\begin{align}
& \hat{\rho} _{U}^{\textrm{exp}} (\mathbf{k},\ellsp , q_{0}) = \frac{1}{\left| \mathbf{k}\right| }b_{1}^{U}+b_{0}^{U}+\mathcal{O}\left( \left| \mathbf{k}\right| \right)  \, .
\end{align}

Defining the velocity is particularly useful as any term with such a factor
vanishes when integrating over $\mathbf{k}$. The velocities can enter also in
the small $\left| \mathbf{k}\right|$ expansion, for instance through
\begin{equation}
  \omega _{k+\ell}=\omega _{\ell}+|\ksp|\, \vdku-|\ksp|^2\frac{(\vdku)^2-1}{2\omega _{\ell}}+
  \mathcal{O}(|\ksp|^3)\,.
\end{equation}

In this appendix, we list the non-vanishing functions $C^{U}$, $A^{U}_{i}$,
$a_{ij}^{U}$ and $b_{i}^{U}$ separately for each diagram~(U). 
\subsection{Diagram (S)}
\begin{figure}[htbp]
\begin{center}
  \includegraphics{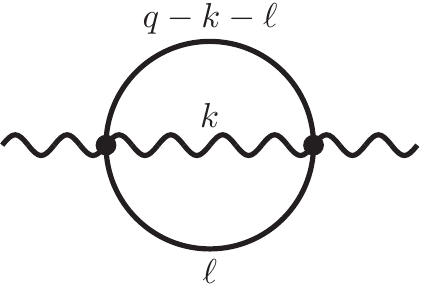}
\end{center}
\caption{Diagram (S)}\label{fig:sdiag}
\end{figure}
First consider (S), whose integrand for the momentum assignment
in~\cref{fig:sdiag} is 
\begin{equation}
  \pi _{S}\left( k,\mathbf{\ell},q_{0}\right)=
\frac{4}{k^{2}\Big( \ell_{0}^{2}+\omega _{\ell}^{2}\Big) \Bigg( \Big( k_{0}+\ell _{0}-q_{0}\Big) ^{2}+\omega _{k+\ell}^{2}\Bigg)  }
  \, .
\end{equation}
The non-vanishing functions entering $\hat{\rho}_{S}$ and
$\hat{\rho}_{S}^{\textrm{exp}}$ are 
\begin{align}
& C^{S} = \frac{-1}{\omega _{\ell}\, \omega _{k+\ell}\Big(\omega _{k+l}+\omega _{l}+\left| \ksp\right| \Big)
  \Bigg(q_0^2+\Big(\omega _{\ell}+\omega _{k+\ell}+|\ksp|\Big)^2\Bigg) } \, , \nonumber \\
& A_{0}^{S} = 1 \, , \nonumber \\
& a_{00}^{S}  = 1\, , \nonumber \\
& b_{1}^{S} = \frac{-1}{2 \,\omega _{\ell}^3 \Big( q_{0}^2+4\omega _{\ell}^2\Big) } \, , \nonumber  \\
& b_{0}^{S}= \frac{q_0^2+12\omega _{\ell}^2-\vdku \, \Big( 3q_0^2+20\omega _{\ell}^2\Big) }{4\omega _{\ell}^4\Big(q_0^2+4\omega _{\ell}^2\Big) ^2} \, .
\end{align}
\subsection{Diagram (T)}
\begin{figure}[htbp]
\begin{center}
  \includegraphics{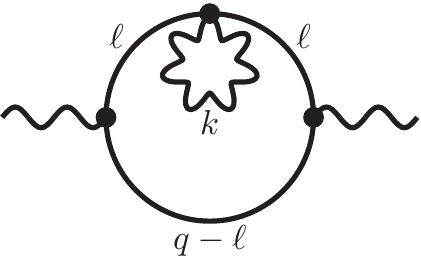}
\end{center}
\caption{Diagram (T)}\label{fig:tdiag}
\end{figure}
Now consider the calculation of diagram (T) with momenta as in~\cref{fig:tdiag}.
The integrand is
\begin{equation}
  \pi _{T}\left( k,\mathbf{\ell},q_{0}\right)= \frac{-16 \, \left| \ellsp \right| ^{2}}{k^{2}\Big( \ell_{0}^{2}+\omega _{\ell}^{2}\Big) ^{2}\Bigg( \Big( \ell _{0}-q_{0}\Big) ^{2} +\omega _{\ell}^{2}\Bigg) }
\, .
\end{equation} 
The non-vanishing functions here are
\begin{align}
& C^{T} =  \frac{\Big( 3q_{0}^{2}+20\omega _{\ell}^{2}\Big)  \left| \ellsp \right| ^{2}}{6\omega _{\ell}^{5}\, \Big( q_{0}^{2}+4\omega _{\ell}^{2}\Big) ^{2} } \, , \nonumber \\
& A_{0}^{T} = 1 \, , \nonumber \\
& a_{00}^{T}  = 1\, , \nonumber \\
& b_{1}^{T}= C^{T} \, .
\end{align}
Note that $ b_{1}^{T}= C^{T}$ since $C^{T}$ cannot be expanded in small $\left|
\mathbf{k} \right|$. Also, since $b_{0}^{T} = 0$ we cannot have any
contributions of order $1/L^{3}$.
\subsection{Diagram (C)}
\begin{figure}[htbp]
\begin{center}
  \includegraphics{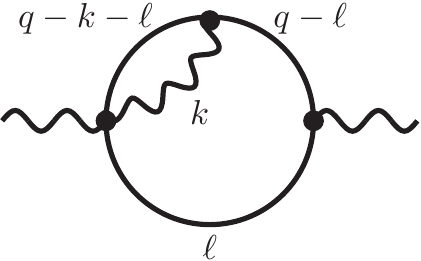}
\end{center}
\caption{Diagram (C)}\label{fig:cdiag}
\end{figure}
For diagram (C) with momenta as in~\cref{fig:cdiag}, the integrand is
\begin{align}
  \pi _{C}\left( k,\mathbf{\ell},q_{0}\right)= \frac{-\left( 8\left| \ellsp \right| ^{2}+4\, \omega _{\ell} \left| \ksp\right| \vdku \right)  }{k^{2}\Big( \ell_{0}^{2}+\omega _{\ell}^{2}\Big) \Bigg( \Big( k_{0}+\ell _{0}-q_{0}\Big) ^{2}+\omega _{k+\ell -q}^{2}\Bigg) \Bigg( \Big( \ell_{0}-q_{0}\Big) ^{2}+\omega _{\ell}^{2}\Bigg)   }
\, .
\end{align} 
This gives
\begin{align}
 C^{C} =&  \frac{\left( 2 \left| \ellsp\right| ^{2}+\omega _{\ell} \left| \ksp \right| \, \vdku\right)
  }{6 \omega_{k+\ell}\,  \omega _{\ell}^3 \Big(q_0^2+4 \omega _{\ell}^2\Big)  \Big(\left| \ksp \right|+\omega_{k+\ell}+\omega _{\ell}\Big)^2 \Bigg(2 \Big(\omega_{k+\ell}+\omega _{\ell}\Big) \left| \ksp \right|+\left| \ksp \right|^2+q_0^2+\Big(\omega_{k+\ell}+\omega _{\ell}\Big)^2\Bigg)}  \, , 
  \nonumber \\
 A_{0}^{C} =& 1 \, , 
 \nonumber \\
 a_{00}^{C}  =& q_0^2 \Big(\omega_{k+\ell}+2 \omega_{\ell}\Big)+\omega_{k+\ell}^3+4 \omega_{k+\ell}^2 \omega_{\ell}+7 \omega_{k+\ell} \omega_{\ell}^2+8 \omega_{\ell}^3\, , \nonumber \\
 a_{01}^{C} =&  q_0^2+3 \omega_{k+\ell}^2+8 \omega_{k+\ell} \omega _{\ell}+7\omega_{\ell}^2\, , 
 \nonumber \\
 a_{02}^{C} = &3 \omega_{k+\ell}+4 \omega_{\ell} \, , 
 \nonumber \\
 a_{03}^{C} =& 1 \, , 
 \nonumber \\
 b_{1}^{C}=&  \frac{\Big(3 q_0^2+20 \omega _{\ell}^2\Big) \left|\ellsp \right|^2}{12 \omega _{\ell}^5 \Big(q_0^2+4 \omega _{\ell}^2\Big)^2}\, , 
 \nonumber \\
 b_{0}^{C}=& \frac{1}{24 \omega _{\ell}^6 \Big(q_0^2+4 \omega _{\ell}^2\Big)^3} \Bigg(\omega _{\ell}^2 \, \vdku \, \Big(3 q_0^4+32 q_0^2 \omega _{\ell}^2+80 \omega _{\ell}^4\Big)
\nonumber \\ &
+2 \left|\ellsp \right|^2 \Big[\vdku \,  \Big(5 q_0^4+54 q_0^2 \omega _{\ell}^2+168 \omega _{\ell}^4\Big)-2 \Big(q_0^4+11 q_0^2 \omega _{\ell}^2+44 \omega _{\ell}^4\Big)\Big]\Bigg) \, .
\end{align}
\subsection{Diagram (E)}
\begin{figure}[htbp]
\begin{center}
  \includegraphics{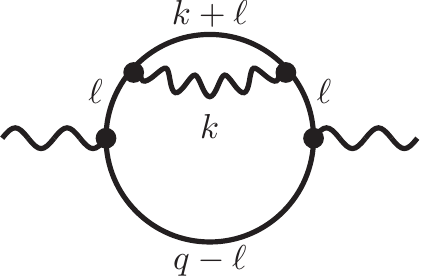}
\end{center}
\caption{Diagram (E)}\label{fig:ediag}
\end{figure}
The integrand for diagram (E), with the momentum assignment in~\cref{fig:ediag},
is
\begin{align}
& \pi _{E}\left( k,\mathbf{\ell},q_{0}\right)=4\frac{\left| \ellsp\right| ^{2}\Bigg( 4\left| \ellsp\right| ^{2} +4l_{0}^{2}+\left| \ksp\right| ^{2}+k_{0}^{2}+4\omega _{\ell}\left| \ksp\right| \vdku +4k_{0}\ell_{0}\Bigg)}{
k^{2}\left( \ell _{0}^{2}+\omega_{\ell}^{2}\right) ^{2}\left( \left( k_{0}+\ell _{0}\right) ^{2}+\omega_{k+\ell}^{2}\right)\left( \left( \ell_{0}-q_{0}\right) ^{2}+\omega_{\ell}^{2}\right)  }\, ,
\end{align}
Here we have
\begin{align}
 C^{E} =&  \frac{\left| \ellsp \right|^2}{96 q_{0}^2 \omega _{k+\ell} \omega _{\ell}^7 \Big(\omega _{k+\ell}+\omega _{\ell}+\left| \ksp \right|\Big)^2}\, , 
  \nonumber \\
 A_{0}^{E} =& -\frac{4 \omega _{\ell}^2}{\omega _{k+\ell}+\omega _{\ell}+\left| \ksp \right|} \, , 
 \nonumber \\
 a_{00}^{E}  =& 4 \Bigg(\omega _{k+\ell} \omega _{\ell}^2 \Big(\omega _{k+\ell}+3 \omega _{\ell}\Big)+\Big(3 \omega _{k+\ell}^2+9 \omega _{k+\ell} \omega _{\ell}+8 \omega _{\ell}^2\Big) \left| \ellsp \right|^2\Bigg) \, , 
 \nonumber \\
 a_{01}^{E} =&3 \omega _{k+\ell}^3+9 \omega _{k+\ell}^2 \omega _{\ell}
   +  13 \omega _{k+\ell} \omega _{\ell}^2+3 \omega _{\ell}^3+12 \Big(2 \omega _{k+\ell}+3 \omega _{\ell}\Big) \left| \ellsp \right|^2
   \nonumber \\ &
   +4 \omega _{\ell} \Big(3 \omega _{k+\ell}^2+9 \omega _{k+\ell} \omega _{\ell}+8 \omega _{\ell}^2\Big) \vdku \, , 
  \nonumber \\
 a_{02}^{E} = &3  \Bigg(4 \left| \ellsp \right|^2+4 \omega _{\ell} \Big(2 \omega _{k+\ell}+3 \omega _{\ell}\Big) \vdku+3 \Big(\omega _{k+\ell}+\omega _{\ell}\Big)^2\Bigg)\, , 
 \nonumber \\
 a_{03}^{E} =&  3 \Bigg( 4 \omega _{\ell} \vdku  
 +3 \Big(\omega _{k+\ell}+\omega _{\ell}\Big) \Bigg)\, , 
 \nonumber \\
  a_{04}^{E} =&3  \, , \nonumber 
\nonumber \\
 A_{1}^{E} =& \frac{16 \omega _{\ell}^4 }{(q_{0}^2+4 \omega _{\ell}^2)^2 (q_{0}^2+(\omega _{k+\ell}+\omega _{\ell})^2+2 (\omega _{k+\ell}+\omega _{\ell}) \left| \ksp \right|+\left| \ksp \right|^2)}\, , 
 \nonumber \\
 a_{10}^{E} =& 4 \Bigg(\Big[q_{0}^4 \Big(\omega _{k+\ell}+2 \omega _{\ell}\Big)
    +4 \omega _{\ell}^2 \Big(3 \omega _{k+\ell}^3
+        12 \omega _{k+\ell}^2 \omega _{\ell}+17 \omega _{k+\ell} \omega _{\ell}^2+8 \omega _{\ell}^3\Big)    
\nonumber \\ &
+q_{0}^2 \Big(\omega _{k+\ell}^3+4 \omega _{k+\ell}^2 \omega _{\ell}+15 \omega _{k+\ell} \omega _{\ell}^2+16 \omega _{\ell}^3\Big)\Big] \left| \ellsp \right|^2
    \nonumber \\ &
    +\omega _{k+\ell} \omega _{\ell}^2 \Big[q_{0}^4
+        4 \omega _{\ell}^2 (\omega _{k+\ell}^2+4 \omega _{k+\ell} \omega _{\ell}+3 \omega _{\ell}^2)+q_{0}^2 (3 \omega _{k+\ell}^2+12 \omega _{k+\ell} \omega _{\ell}+13 \omega _{\ell}^2)\Big]
    \Bigg)\, , 
 \nonumber \\
a_{11}^{E} =&  q_{0}^4 \omega _{k+\ell}^2+q_{0}^2 \omega _{k+\ell}^4+
    2 q_{0}^4 \omega _{k+\ell} \omega _{\ell} 
    +4 q_{0}^2 \omega _{k+\ell}^3 \omega _{\ell} 
 +q_{0}^4 \omega _{\ell}^2+42 q_{0}^2 \omega _{k+\ell}^2 \omega _{\ell}^2+76 q_{0}^2 \omega _{k+\ell} \omega _{\ell}^3
 \nonumber \\ &  
 +12 \omega _{k+\ell}^4 \omega _{\ell}^2+48 \omega _{k+\ell}^3 \omega _{\ell}^3+13 q_{0}^2 \omega _{\ell}^4
+     104 \omega _{k+\ell}^2 \omega _{\ell}^4+112 \omega _{k+\ell} \omega _{\ell}^5  
    +12 \omega _{\ell}^6
 \nonumber \\ &    
    +4 \Bigg(q_{0}^4+q_{0}^2 \Big(3 \omega _{k+\ell}^2+8 \omega _{k+\ell} \omega _{\ell}+15 \omega _{\ell}^2\Big)+4 \omega _{\ell}^2 \Big(9 \omega _{k+\ell}^2+24 \omega _{k+\ell} \omega _{\ell}
    +
    17 \omega _{\ell}^2\Big)\Bigg) \left| \ellsp \right|^2
\nonumber \\ &     
    +   
    4 \omega _{\ell} \Bigg(q_{0}^4 \Big(\omega _{k+\ell}+2 \omega _{\ell}\Big)+4 \omega _{\ell}^2 \Big(3 \omega _{k+\ell}^3+12 \omega _{k+\ell}^2 \omega _{\ell} +17 \omega _{k+\ell} \omega _{\ell}^2+8 \omega _{\ell}^3\Big)
      \nonumber \\ & 
    +q_{0}^2 \Big(\omega _{k+\ell}^3
    +4 \omega _{k+\ell}^2 \omega _{\ell}
+        15 \omega _{k+\ell} \omega _{\ell}^2+16 \omega _{\ell}^3\Big)\Bigg) \vdku \, , 
 \nonumber \\
a_{12}^{E} =& 2 \Bigg(q_{0}^4 \omega _{k+\ell}  
    +2 q_{0}^2 \omega _{k+\ell}^3+q_{0}^4 \omega _{\ell}+6 q_{0}^2 \omega _{k+\ell}^2 \omega _{\ell}+24 q_{0}^2 \omega _{k+\ell} \omega _{\ell}^2
+       24 \omega _{k+\ell}^3 \omega _{\ell}^2+14 q_{0}^2 \omega _{\ell}^3
          \nonumber \\ &
    +72 \omega _{k+\ell}^2 \omega _{\ell}^3+80 \omega _{k+\ell} \omega _{\ell}^4 
    +24 \omega _{\ell}^5+2 \Big(3 \omega _{k+\ell}+4 \omega _{\ell}\Big) \Big(q_{0}^2+12 \omega _{\ell}^2\Big) \left| \ellsp \right|^2+2 \omega _{\ell} \Big[q_{0}^4
+    
\nonumber \\ &  
    q_{0}^2 \Big(3 \omega _{k+\ell}^2+8 \omega _{k+\ell} \omega _{\ell}+15 \omega _{\ell}^2\Big) 
    +4 \omega _{\ell}^2 \Big(9 \omega _{k+\ell}^2+24 \omega _{k+\ell} \omega _{\ell}+17 \omega _{\ell}^2\Big)\Big] \vdku\Bigg)\, , 
 \nonumber \\
  a_{13}^{E} =& \Big( q_{0}^2+12 \omega _{\ell}^2 \Big) \Bigg(q_{0}^2 
    +6 \omega _{k+\ell}^2+12 \omega _{k+\ell} \omega _{\ell}+6 \omega _{\ell}^2
    +4 \left| \ellsp \right|^2+4 \omega _{\ell} \Big(3 \omega _{k+\ell}+4 \omega _{\ell}\Big) \vdku\Bigg)\, , 
 \nonumber \\
  a_{14}^{E} =& 4 \Big( q_{0}^2+12 \omega _{\ell}^2\Big) \Big(\omega _{k+\ell}+\omega _{\ell}+\omega _{\ell} \vdku\Big)\, , 
 \nonumber \\
  a_{15}^{E} =& q_{0}^2+12 \omega _{\ell}^2\, , 
 \nonumber \\
 b_{1}^{E}=&  \frac{-\left| \ellsp \right|^2 \Bigg(\omega _{\ell}^2 \Big(q_{0}^4+10 q_{0}^2 \omega _{\ell}^2-8 \omega _{\ell}^4\Big)
+\Big(5 q_{0}^4+54 q_{0}^2 \omega _{\ell}^2+168 \omega _{\ell}^4\Big) \left| \ellsp \right|^2\Bigg)}{12 \omega _{\ell}^7 \Big(q_{0}^2+4 \omega _{\ell}^2\Big)^3 }\, , 
 \nonumber \\
 b_{0}^{E}=& \frac{\left| \ellsp \right|^2}{48 \omega _{\ell}^8 \Big(q_{0}^2+4 \omega _{\ell}^2\Big)^4}
 \Bigg( -\omega _{\ell}^2 \Big[\Big(25 q_{0}^6+368 q_{0}^4 \omega _{\ell}^2+1872 q_{0}^2 \omega _{\ell}^4+2688 \omega _{\ell}^6\Big) \vdku 
\nonumber \\ & +q_{0}^6+16 q_{0}^4 \omega _{\ell}^2
+  80 q_{0}^2 \omega _{\ell}^4+640 \omega _{\ell}^6\Big]
\nonumber \\ &+\left| \ellsp \right|^2 \Big[-\Big(35 q_{0}^6+520 q_{0}^4 \omega _{\ell}^2+2736 q_{0}^2 \omega _{\ell}^4+5376 \omega _{\ell}^6\Big) \vdku +15 q_{0}^6
\nonumber \\ &
+224 q_{0}^4 \omega _{\ell}^2
+  1200 q_{0}^2 \omega _{\ell}^4+2688 \omega _{\ell}^6\Big]
 \Bigg) \, .
\end{align}
\subsection{Diagram (X)}
\begin{figure}[htbp]
\begin{center}
  \includegraphics{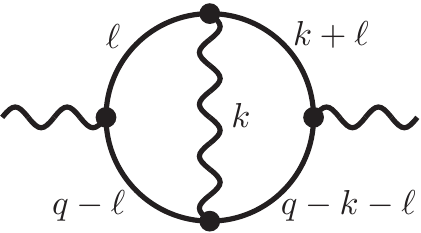}
\end{center}
\caption{Diagram (X)}\label{fig:xdiag}
\end{figure}
Assigning momenta as in~\cref{fig:xdiag}, the integrand of diagram (X) is
\begin{align}
&\pi _{X}\left( k,\ell,q_{0}\right) =
\Bigg( \Big(k_{0} + 2 \ell _{0}\Big) \Big[-k_{0} +2 \Big(-\ell _{0} + 
           q_{0}\Big)\Big] - \left|\mathbf{k}\right|^2 - 
     4 \left|\ellsp\right|^2 - 
     4 \omega _{\ell} \left|\mathbf{k}\right| 
\vdku\Bigg)
\nonumber \\
&\times 
\frac{-4 \Big(\left| \ellsp\right| ^2 + \omega _{\ell} \left|\mathbf{k}\right| \vdku\Big) }{ \Bigg( \Big(k_{0} +   \ell _{0}\Big)^2 + \omega _{k+\ell}^2\Bigg) \Bigg(\Big(-k_{0} - \ell _{0} + 
       q_{0}\Big)^2 + \omega _{k+\ell}^2\Bigg) \Big( \ell _{0}^2 + \omega _{\ell}^2\Big) \Bigg(\Big(-\ell _{0} + 
       q_{0}\Big)^2 + \omega _{\ell}^2\Bigg) k^{2}}\, .
 \end{align}
The non-vanishing functions are now
\begin{align}
C^{X}= &  \frac{\left| \ellsp \right|^2+\omega _{\ell} \left| \ksp \right| \vdku}{12 q_{0}^2 \omega _{k+\ell}^3 \omega _{\ell}^3 \Big(\omega _{k+\ell}+\omega _{\ell}+\left| \ksp \right|\Big)}  \, ,
\nonumber \\
A_{0}^{X}= & \frac{-1}{\Big(\omega _{k+\ell}+\omega _{\ell}+\left| \ksp \right|\Big) ^2}\, ,
\nonumber \\
a_{00}^{X} = & 8 \Bigg(\omega _{k+\ell}^2 \omega _{\ell}^2+\Big(\omega _{k+\ell}^2+3 \omega _{k+\ell} \omega _{\ell}+\omega _{\ell}^2\Big) \left| \ellsp \right|^2\Bigg)\, ,
\nonumber \\
a_{01}^{X} = & \Big(\omega _{k+\ell}+\omega _{\ell}\Big)^3+12 \Big(\omega _{k+\ell}+\omega _{\ell}\Big) \left| \ellsp \right|^2+8 \omega _{\ell} \Big(\omega _{k+\ell}^2+3 \omega _{k+\ell} \omega _{\ell}+\omega _{\ell}^2\Big) \vdku \, ,
\nonumber \\
a_{02}^{X} = &4 \left| \ellsp \right|^2
+   3 \Big(\omega _{k+\ell}+\omega _{\ell}\Big) \Big(\omega _{k+\ell}+\omega _{\ell}+4 \omega _{\ell} \vdku\Big) \, ,
\nonumber \\
a_{03}^{X} = &4 \omega _{\ell} \vdku 
   + 3 \Big(\omega _{k+\ell}+\omega _{\ell}\Big) \, ,
\nonumber \\
a_{04}^{X} = &1 \, ,
\nonumber \\
A_{1}^{X}= &  \frac{16 \omega _{k+\ell}^2 \omega _{\ell}^2}{\Big(q_{0}^2+4 \omega _{k+\ell}^2\Big) \Big(q_{0}^2+4 \omega _{\ell}^2\Big) \Bigg(q_{0}^2+\Big(\omega _{k+\ell}+\omega _{\ell}\Big)^2+2 \Big(\omega _{k+\ell}+\omega _{\ell}\Big) \left| \ksp \right|+\left| \ksp \right|^2\Bigg)}\, ,
\nonumber \\
a_{10}^{X} = & 2 \omega _{k+\ell} \omega _{\ell} \Big(-3 q_0^2
+4 \omega _{k+\ell} \omega _{\ell}\Big) +2 \Bigg( q_0^2+4 \Big(\omega _{k+\ell}^2+3 \omega _{k+\ell} \omega _{\ell}+\omega _{\ell}^2\Big)\Bigg) \left| \ellsp \right|^2\, ,
\nonumber \\
a_{11}^{X} = & 12 \Big(\omega _{k+\ell}+\omega _{\ell}\Big) \left| \ellsp \right|^2+\Big(\omega _{k+\ell}+\omega _{\ell}\Big) \Bigg(-2 q_0^2+\Big(\omega _{k+\ell}+\omega _{\ell}\Big)^2\Bigg)
    \nonumber \\ &
    +    
    2 \omega _{\ell} \Bigg(q_0^2+4 \Big(\omega _{k+\ell}^2+3 \omega _{k+\ell} \omega _{\ell}+\omega _{\ell}^2\Big)\Bigg) \vdku\, ,
\nonumber \\
a_{12}^{X} = & 3 \Big( \omega _{k+\ell}+\omega _{\ell}\Big) \Big(\omega _{k+\ell}
    +\omega _{\ell}+4 \omega _{\ell} \vdku \Big)+4 \left| \ellsp \right|^2 \, ,
\nonumber \\
a_{13}^{X} = & 3 \Big(\omega _{k+\ell}+\omega _{\ell}\Big)+4 \omega _{\ell} \vdku\, ,
\nonumber \\
a_{14}^{X} = &1 \, ,
\nonumber \\
b_{1}^{X} = & \frac{-\left| \ellsp \right|^2}{12 \omega _{\ell}^7 (q_{0}^2+4 \omega _{\ell}^2)^3 }
 \Bigg(\omega _{\ell}^2 \Big(q_{0}^4+12 q_{0}^2 \omega _{\ell}^2+96 \omega _{\ell}^4\Big)+\Big(5 q_{0}^4+60 q_{0}^2 \omega _{\ell}^2+224 \omega _{\ell}^4\Big) \left| \ellsp \right|^2\Bigg)\, , 
\nonumber \\
b_{0}^{X} = &  \frac{-1}{24 \omega _{\ell}^8 (q_{0}^2+4 \omega _{\ell}^2)^4}
 \Bigg(2 \omega _{\ell}^4 \Big(q_{0}^6+16 q_{0}^4 \omega _{\ell}^2+144 q_{0}^2 \omega _{\ell}^4+384 \omega _{\ell}^6\Big)
   \vdku
  -\omega _{\ell}^2 \left| \ellsp \right|^2 \Big[-q_{0}^6
  \nonumber \\ & 
  -16 q_{0}^4 \omega _{\ell}^2 
  -  80 q_{0}^2 \omega _{\ell}^4-640 \omega _{\ell}^6+\Big(25 q_{0}^6+400 q_{0}^4 \omega _{\ell}^2+2384 q_{0}^2 \omega _{\ell}^4+6272 \omega _{\ell}^6\Big) \vdku\Big]
\nonumber \\ &   
 +\left| \ellsp \right|^4 \Big[\Big(35 q_{0}^6+560 q_{0}^4 \omega _{\ell}^2+3312 q_{0}^2 \omega _{\ell}^4+8064 \omega _{\ell}^6\Big) \vdku-9 q_{0}^6-144 q_{0}^4 \omega _{\ell}^2
\nonumber \\ &  
  -  848 q_{0}^2 \omega _{\ell}^4-2176 \omega _{\ell}^6\Big]\Bigg)\, .
\end{align}

\pagebreak
\bibliography{hvpqedfv}
\end{document}